\documentclass[manuscript]{aastex}

\newcommand{\etal}{et al.}
\font\eightrm=cmr8

\newbox\grsign \setbox\grsign=\hbox{$>$} \newdimen\grdimen \grdimen=\ht\grsign
\newbox\simlessbox \newbox\simgreatbox
\setbox\simgreatbox=\hbox{\raise.5ex\hbox{$>$}\llap
     {\lower.5ex\hbox{$\sim$}}}\ht1=\grdimen\dp1=0pt
\setbox\simlessbox=\hbox{\raise.5ex\hbox{$<$}\llap
     {\lower.5ex\hbox{$\sim$}}}\ht2=\grdimen\dp2=0pt
\def\simgreater{\mathrel{\copy\simgreatbox}}
\def\simless{\mathrel{\copy\simlessbox}}
\newbox\simppropto
\setbox\simppropto=\hbox{\raise.5ex\hbox{$\sim$}\llap
     {\lower.5ex\hbox{$\propto$}}}\ht2=\grdimen\dp2=0pt

\slugcomment{To appear in The Astronomical Journal}

\shorttitle{M~31 \& MW Globular Clusters}
\shortauthors{Schiavon et al.}

\begin{document}


\title{Star Clusters in M~31. IV. A Comparative
Analysis of Absorption Line Indices in Old M~31 and Milky Way Clusters}


\author{Ricardo P. Schiavon}
\affil{Gemini Observatory, 670 N. A'Ohoku Place, Hilo, HI 96720}
\email{rschiavon@gemini.edu}

\author{Nelson Caldwell}
\affil{Smithsonian Astrophysical Observatory, 60 Garden Street, Cambridge, MA 02138, USA}
\email{caldwell@cfa.harvard.edu}

\author{Heather Morrison and Paul Harding}
\affil{Department of Astronomy, Case Western Reserve University, Cleveland, OH 44106-7215, USA}
\email{paul.harding@case.edu, heather@vegemite.case.edu}


\author{St\'ephane Courteau}
\affil{Department of Physics, Engineering Physics \& Astronomy, Queen's University,
	   Kingston, ON, Canada K7L 3N6}
\email{<courteau@astro.queensu.ca>}

\author{Lauren A. MacArthur}
\affil{Herzberg Institute of Astrophysics, National Research Council of
	Canada/ University of Victoria, Victoria, B.C., V9E 2E7 Canada}
\email{Lauren.MacArthur@nrc-cnrc.gc.ca}

\and 

\author{Genevieve J. Graves}
\affil{Department of Astronomy, University of California, Berkeley, CA
	94720}
\email{graves@astro.berkeley.edu}



\begin{abstract}
We present absorption line indices measured in the integrated spectra
of globular clusters both from the Galaxy and from M~31.  Our samples
include 41 Galactic globular clusters, and more than 300 clusters
in M~31.  The conversion of instrumental equivalent widths into the
Lick system is described, and zero-point uncertainties are provided.
Comparison of line indices of old M~31 clusters and Galactic
globular clusters suggests an absence of important differences in
chemical composition between the two cluster systems.  In particular,
CN indices in the spectra of M~31 and Galactic clusters are essentially
consistent with each other, in disagreement with several previous
works.  We reanalyze some of the previous data, and conclude that
reported CN differences between M~31 and Galactic clusters were
mostly due to data calibration uncertainties.  Our data support the
conclusion that the chemical compositions of Milky Way and M~31
globular clusters are not substantially different, and that there
is no need to resort to enhanced nitrogen abundances to account for
the optical spectra of M~31 globular clusters.
\end{abstract}





\section{Introduction}

The history of star formation and chemical enrichment of galaxies
is encoded in the ages and chemical compositions of their stellar
populations.  In particular, powerful insights on the processes
leading to the assembly of the Galactic halo are gained by studies
of the chemical abundances of their constituent populations of field
stars and globular clusters (GCs).  It is only natural to extend
such studies to the nearest giant spiral galaxy, M~31.  While
detailed abundances of individual stars in the Galactic halo, disk,
and bulge, have been obtained and extensively analyzed in the past
several decades, similar studies of comparable samples of M~31 halo
stars await the advent of 20-30~m class telescopes, equipped with
efficient high-resolution spectrographs.  In the meantime, integrated
light studies of M~31 GCs have historically provided valuable
qualitative information about the chemical composition of the halo
of M~31, leading to important clues on its formation history.  As
stellar population synthesis models and instruments both grow in
sophistication, quantitative information on elemental abundances
from integrated-light studies of M~31 GCs is also becoming available
\citep[e.g.,][]{co09,ca10}.

The history of this field has been punctuated by heroic observational
efforts, based on optical spectra painstakingly collected with,
according to today's standards, relatively small telescopes, rather
inefficient spectrographs, and low-quantum efficiency (and often
difficult to calibrate) detectors \citep[e.g.,][]{bu84,bh90}.  The
main results emerging from these efforts are: 1) M~31 GCs
are on average slightly more metal rich than their Galactic
counterparts, while spanning roughly the same range of metallicities
\citep{bh91}; 2) M~31 GCs are $\alpha$-enhanced, just like
those in the MW \citep[e.g.,][]{pu05}; 3) M~31 GCs are CN-enhanced
compared to MW GCs \citep{bu84,bh90,be04,pu05}.  In particular,
\cite{bu84} found that the CN band at $\lambda$ 4170 ${\rm\AA}$ is
stronger by $\sim$ 0.05 mag in the spectra of moderately metal-poor
M31 GCs than in those of their MW counterparts of same metallicity;
and 4) M~31 GCs are possibly younger/older than those in the
MW.  \cite{bu84} found that $H\beta$ is stronger in the spectra of
M~31 GCs than in those of their MW counterparts by about 0.5
${\rm\AA}$, indicating a younger age or differences in horizontal
branch morphology.  The latter two results are contradicted by the
findings presented in this paper.  While \cite{ca10} discuss the
Balmer line strengths in the spectra of the two GC systems,
comparing their ages, CN is the main focus of this paper.

Those significant conclusions may teach us about important aspects
of the formation of the two galaxies.  For instance, the comparable
mean metallicities may be indicative of similar overall chemical
enrichment, suggesting that the early star formation efficiency has
been similar in the two systems---or perhaps in the sub-components
that eventually assembled to form them.  The similar $\alpha$-enhancement,
assessed by measurements of Mg and Fe-sensitive absorption lines,
suggests that either the time scale for star formation, the IMF,
or a combination thereof, were similar in both galaxies.  The
difference in CN strength, which has been ascribed to a difference
in nitrogen abundance, is difficult to interpret, owing mostly to
uncertainties in the models for the nucleosynthesis of that element.
The issue is further complicated by the fact that CNO elements are
seen to present strong star-to-star variations in Galactic GCs
\citep[e.g.,][and references therein]{gr04}, which may be associated
with the presence of multiple stellar populations in those GCs
\citep[][and references therein]{cs10,pi09,ms09,can98}, and by
inference, in their M~31 counterparts.  Regardless, any scenario
for the formation of the MW and M~31 haloes, and their GC systems,
will be challenged by the large nitrogen abundance differences
between systems that look otherwise very similar.


This is the fourth of a series of papers dedicated to analyzing the
kinematics and chemistry of a large sample of M~31 GCs, based on
high-quality integrated spectra for several hundred M~31 clusters,
obtained with MMT/Hectospec.  In Paper I \citep{ca09a} we characterized
the population of young ($\simless$ 2 Gyr) M~31 clusters in terms
of their ages, metallicities, masses, and kinematics.  In Paper II
\citep{ca10}, the ages and metallicities of M~31 old GCs were
studied.  In particular, we found no differences between the ages
of the two old GC systems, in disagreement with previous claims.
Moreover, we found that the M~31 GC system does not have a bimodal
metallicity distribution, in agreement with recent findings
\citep{yo06}.  In Paper III \citep{mo10}, we suggest that the old
bulge GCs in M~31 are characterized by a bar-like kinematics.

In this paper, we present absorption line indices measured in
integrated spectra of a large sample of M~31 and MW GCs.
The data for M~31 GCs come from spectra described by \cite{ca09a}, and
those for MW GCs are based on the integrated GC spectra
from \cite{s05}.  While the focus of this paper is on the comparison
of CN strengths between the two GC systems, \cite{s11} will
present an analysis of abundance ratios, based on application of
stellar population synthesis models to these index measurements.
This paper is organized as follows: in Section~\ref{data}, the
measurement of Lick absorption line indices is described, and a
comparison between CN indices in M~31 and MW GCs is presented
in Section~\ref{analysis}.  Our conclusions are summarized in
Section~\ref{epilogue}.

\section{Data} \label{data}

A detailed description of our data, including sample selection,
observation and reduction procedures, is presented in \cite{s05}
and \cite{ca09a} for the MW and M~31 GCs, respectively.  Various
aspects that are relevant to this work are summarized below.

The M~31 GC spectra were obtained using the multi-fiber spectrograph
Hectospec \citep{fa05}, attached to the 6.5~m Multiple Mirror
Telescope (MMT).  The projected diameter of a Hectospec fiber on
the sky is 1$\arcsec$.5, and the spectral coverage of the resulting
1-D, wavelength-calibrated spectra, is 3650--9200 ${\rm\AA}$.
Spectral dispersion, and resolution are 1.2 ${\rm \AA}$ pixel$^{-1}$
and 5${\rm\AA}$, respectively, and the median S/N of the spectra
is 75 pixel$^{-1}$.  The normal operating procedure with Hectospec,
and other multi-fiber spectrographs, is to assign a number of fibers
to blank sky areas in the focal plane, and then combine those in
some fashion to allow sky subtraction  of the target spectra.  These
methods are satisfactory for our outer M~31 fields, but not for the
bulge areas where the local background is high relative to the
cluster targets (and where the most metal-rich clusters reside,
along with many metal-poor clusters).  As described in Paper II,
for the bulge clusters only sky spectra far from the bulge of M31
were used.  A separate offset exposure for such  fields, taken
concurrently and about 5$\arcsec$ offset from the targets, was reduced
in a similar way (so that contemporaneous sky subtraction was
performed for on- and off-target exposures), and then these off-target
local background spectra were subtracted from the on-target.
Relative flux calibration, aimed at removing the signatures of
instrumental and atmospheric transmission, was achieved using
observations of flux standards.

The MW GC spectra were collected with the CTIO Blanco 4~m telescope,
equipped with the Ritchey-Chr\'etien spectrograph, mounted at the
telescope's Cassegrain focus.  Given the extended nature of Galactic
GCs, observations were executed by drift scanning the targets with
a 5$\arcmin$.5-long slit, over the range of $\pm$ one GC core radius,
$r_c$, taken from \cite{ha96}.  Additional exposures in areas
surrounding the target GCs were obtained for background-subtraction
purposes.  One-dimensional spectra were extracted by coadding the
columns contained within a $\pm\,\sim\,1\,r_c$ spatial window
centered on the peak of each GC's light profile.  Therefore, for
most GCs, the 1-D spectra sample a core radius-sized square spatial
region \citep[but see ][for exceptions]{s05}.  No significant
variations in Lick index measurements were found between spectra
that sample different spatial regions, for any of the 9 GCs for
which such spectra were available.  The spectra were wavelength-calibrated
in the usual fashion.  The resulting 1D wavelength-calibrated spectra
cover the region between 3360 and 6430 ${\rm\AA}$, with a spectral
dispersion of 1 ${\rm\AA}$ pixel$^{-1}$ and a resolution of $\sim$
3.1 ${\rm\AA}$.  Relative flux calibration was achieved in the usual
fashion, using observations of spectrophotometric standards.

\subsection{Absorption line indices} \label{absind}

Even though the two sets of GC spectra were obtained with different
instruments, and following rather different observational procedures,
the measurement of line indices in the 1D spectra and their conversion
from instrumental to standard system follows the same well established
recipes that are only briefly described in this Section.  All line
indices discussed in this paper were measured with the {\tt lick\_ew}
code, which is part of the {\tt EZ\_Ages}\footnote{\eightrm
http://www.ucolick.org/$\sim$graves/ez\_ages.html} package \citep{gs08}.
 Cluster spectra were first smoothed from their original resolution
to the resolution of the Lick index, system, given in Table 1 of
\cite{s07} for each of the indices.  The original resolution was
5${\rm\AA}$ in the case of the M~31 GCs, and that given in equation
(1) of \cite{s05}.  Instrumental index measurements are based on
the definitions provided by \cite{wo94} and \cite{wo97}.  Conversion
between instrumental and standard Lick indices was achieved in the
usual fashion, through comparison with measurements taken in spectra
of standard stars.  Observation of Lick standards were obtained
using the same instrumental setup used for GC observations, and the
spectra reduced following the same procedures as adopted for GC
spectra.  Likewise, the spectra of Lick standards were smoothed to
the Lick resolution, and instrumental indices measured adopting the
same definitions as above.  For the M~31 Hectospec data, the six
standards were observed through single fibers, though each star was
observed through a different fiber.  No significant star-to-star
variation is seen in the residuals shown in Figures~\ref{zm31a} and
\ref{zm31b}, so we conclude that different fiber-to-fiber throughput
variations contribute negligible uncertainties to our index
measurements.

We note that the Tololo spectra are affected by a bad CCD column
that runs across the spectra at a wavelength placed within the
passband of the Lick CN indices.  We measured the effect of that
bad column on our index measurements and found them to be affected
at the $\sim$ 0.015 mag level in the worst cases.  Moreover, the
effect varies in sign and intensity from GC to GC , and thus should
not introduce any significant systematic effect in our measurements.

An important aspect of our conversion to the Lick system is the
adoption of standard index values from \cite{s07}, rather than those
of \cite{wo94}.  We proceeded this way because of the higher accuracy
of the \cite{s07} indices.  Because they are based on flux-calibrated
CCD spectra, they are free from the uncertainties associated with
difficulties in calibrating data obtained with the Lick Image
Dissector Scanner, upon which the \cite{wo94} indices are based.
For more details, see discussion in \cite{s07}, and particularly
Figures 1 and 2 of that paper.

\begin{figure}
\includegraphics[angle=0,scale=0.8]{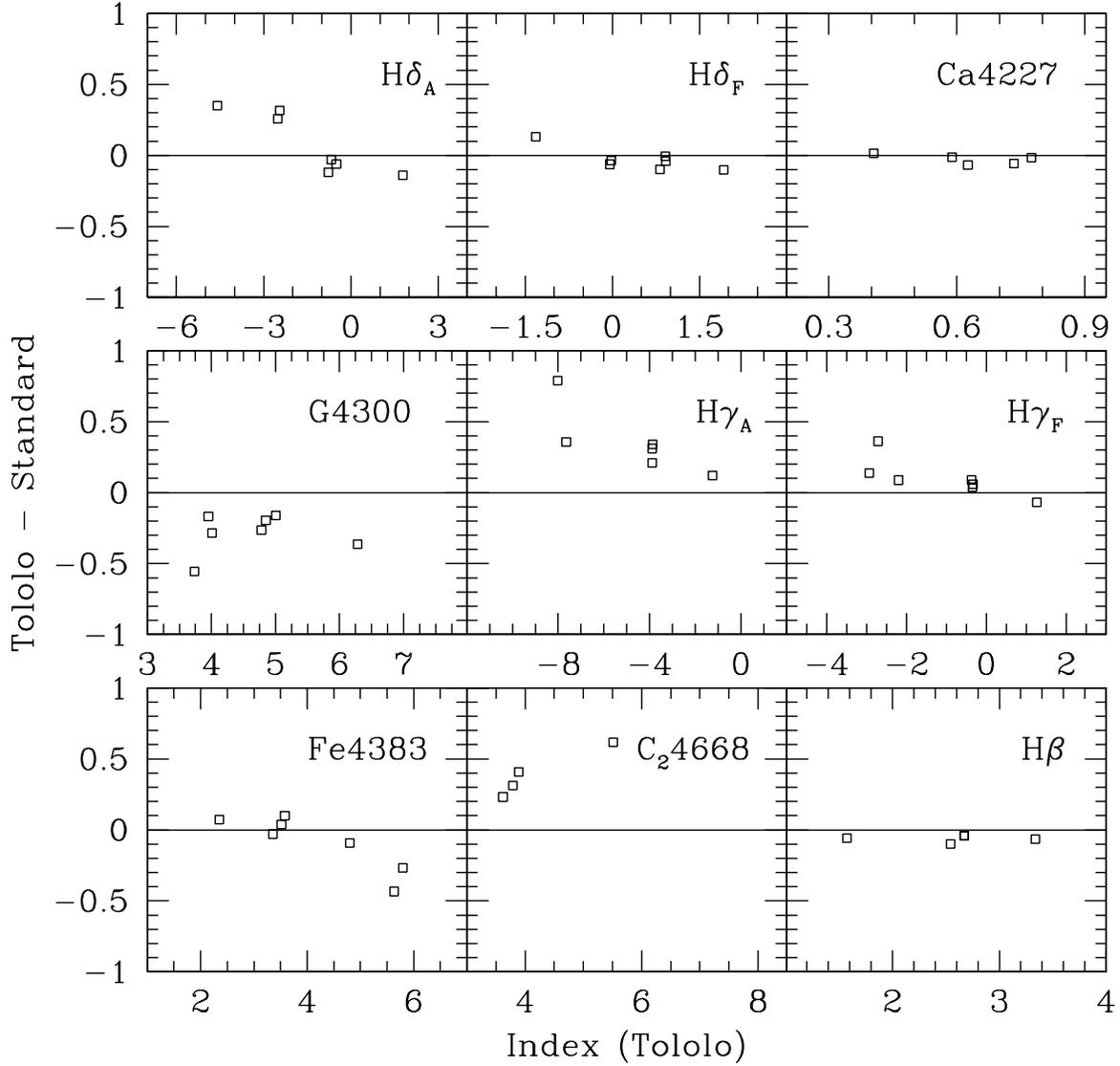}
\caption{Calibration of instrumental pseudo-equivalent widths into
the Lick system for the Tololo integrated spectra of Galactic GCs
from \cite{s05}, using observations of Lick standards.  Zero points
were calculated by averaging the residuals for all indices, excluding
very strong deviants.  Standard Lick indices come from \cite{s07}.
Residuals are plotted as a function of instrumental index strength.
Formal error bars based on photon noise, readout noise, and
sky-subtraction errors are smaller than the symbol sizes and are
not plotted.  The few standard star observations yield somewhat
uncertain zero points for some indices.  See discussion in the text.
}
\label{zeroa}
\end{figure}

\begin{figure}
\includegraphics[angle=0,scale=0.8]{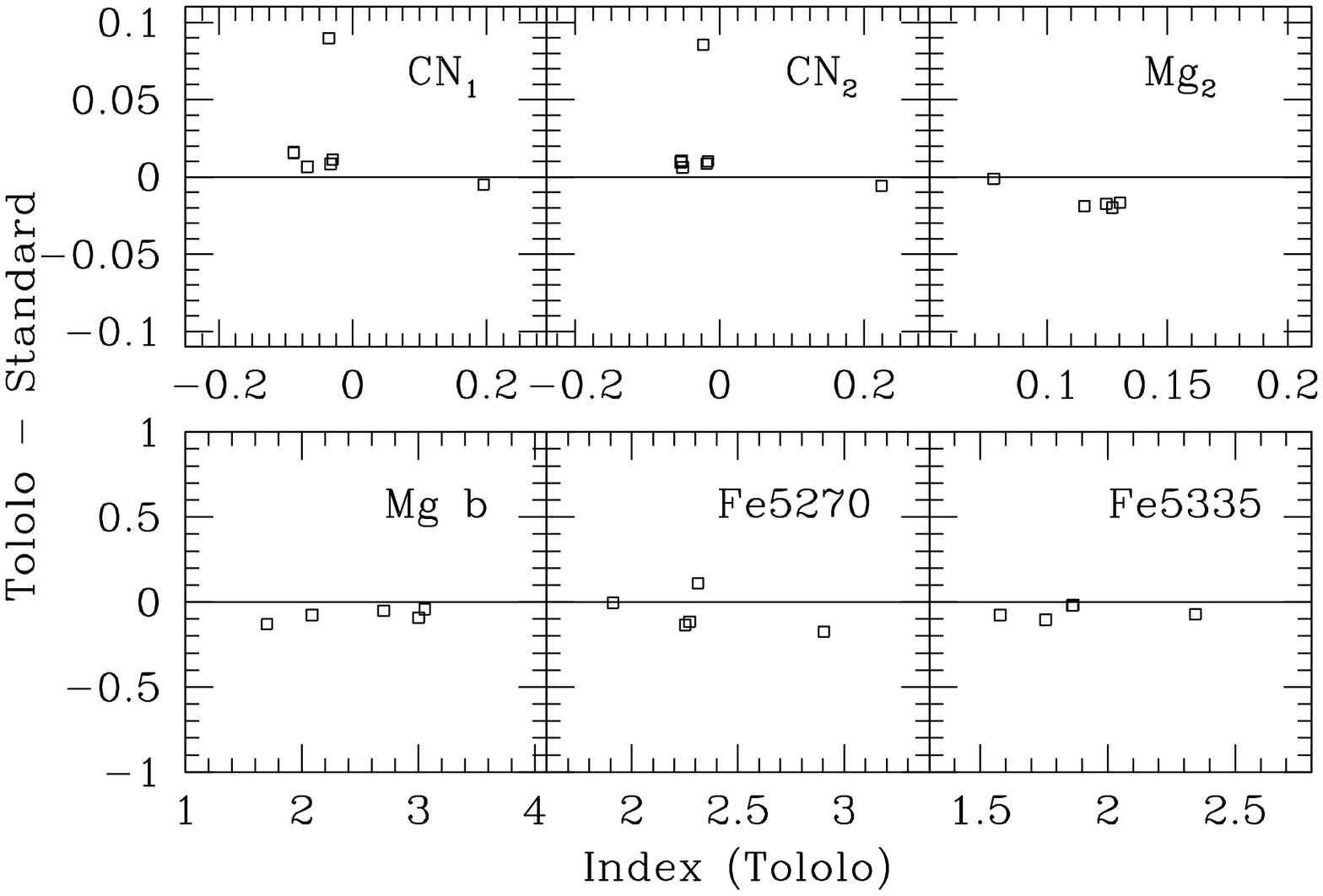}
\caption{Figure~\ref{zeroa} continued.  Note that, except for a
single very deviant measurement, residuals for CN indices are very
small---in fact, much smaller than the $\sim$ 0.05 mag CN differences
measured by \cite{bu84} between MW and M~31 GCs.  The very deviant
point is due to star HD 78418, which is a spectroscopic binary.  It is not
clear why it is not very deviant in other indices as well.  Comparison of
the Tololo spectrum for this star with that from \cite{jo99} suggests that
there may be problems with flux calibration of the former, in the region of
the CN indices.
}
\label{zerob}
\end{figure}

Instrumental and standard Lick indices are compared in Figures~\ref{zeroa}
and \ref{zerob} for the Tololo data, and in Figures ~\ref{zm31a}
and \ref{zm31b} for the Hectospec data, where residuals are plotted
as a function of index strength.  Error bars were calculated based
on propagation of photon noise, readout noise, and sky-subtraction
errors into index measurements, following \cite{ca98}.  In the case
of the Tololo data, they are smaller than the symbol sizes, and are
thus not plotted.  The IDs of the standard stars observed, as well
as the instrumental measurements are provided in Tables~\ref{std_m31a}
through \ref{std_mwb}.  The standard values were taken from \cite{s07},
for the indices modeled in that paper.  The residuals are in general
small, and so are the zero point corrections for all indices in
both sets of observations, which is a result of the data being based
on flux calibrated, high-S/N spectra, as discussed by \cite{s07}.
We note that, due mostly to unfavorable weather, we unfortunately
were unable to collect data for a large number of Lick standards,
particularly during the observing run in Cerro Tololo, where the
Galactic GCs were observed.  Therefore, as seen in those Figures,
the conversion to the standard Lick system is sometimes based on a
handful of data points, in most cases spanning an insufficient range
of index strengths, resulting in somewhat uncertain, though small,
zero point corrections.  Most importantly, for the purposes of this
paper, we note that the residuals of the CN indices in both data
sets are very small---indeed, much smaller than the 0.05 mag CN
difference found by \cite{bu84} between the M~31 and MW GC systems.

For some indices, the data suggest the presence of a correlation
of the residuals with index strength.  That is the case of $C_24668$
for the M~31 sample, and $H\delta_A$, $H\gamma_A$, $H\gamma_F$, and
Fe4383, in the case of the MW sample.  Despite the evidence for
such trends, we adopt single zero-point conversions for such indices,
because the range spanned by the residual variations is relatively
small.  For instance, according to the \cite{s07} models, the
systematics introduced in [C/Fe] determinations by adopting a
constant zero-point conversion to $C_24668$ is smaller than $\sim$
0.1 dex.  By the same token, uncertainties in [Fe/H] due to zero
point uncertainties in Fe4383 are smaller than 0.1 dex, and those
in the Balmer lines translate into a mere $\sim$ 1 Gyr uncertainty
in age, for old systems.


The final zero points are thus estimated as the mean residual,
excluding strongly deviant points.  The paucity of data points makes
it difficult to estimate the zero point uncertainties, and to be
as fair as possible, we define these as 1/2 of the dynamic range
of the residuals, or:

\begin{equation} \label{zpt}
\Delta Zero\,\,Point = {Max(Residual) - Min(Residual) \over 2},
\end{equation}

\noindent where strongly deviant points were removed from the
statistics.  This may be a fair representation of the uncertainties
in indices that present a large scatter of the residuals (e.g.,
G4300) or showing an apparent trend of residuals with index strength
($H\delta_A$, $H\gamma_A$, $H\gamma_F$, and Fe4383), but it possibly
underestimates the uncertainties for indices where little variation
of the residuals is seen.  However, the latter is presumably due
to the indices being robust to instrumental effects, as is probably
the case of $H\beta$ and redder indices---see discussion and plots
in \cite{s04a} and \cite{s07}.  The final zero point conversions
to the Lick system from \cite{s07} and associated uncertainties are
provided in Tables~\ref{zeromw} and ~\ref{zerom31}.

Finally, we decided to include, for completeness, indices that were
not modeled by \cite{s07} \footnote{\eightrm Recall that the
\cite{s07} models are based on the \cite{jo99} library, which does
not cover the entire spectral range spanned by the Lick system},
and for those they come from unpublished measurements taken on the
spectra of the Indo-US library \citep{va04}, convolved to the Lick/IDS
resolution \citep[see details in][]{s07} \footnote{\eightrm These
index measurements can be obtained under request to the authors}.
Conversion of the instrumental indices into the Indo-US system is
provided in Tables~\ref{convmw} and \ref{convm31} for MW and M~31 
GCs, respectively.  Conversion between Indo-US values and the
standard Lick/IDS system of \cite{wo94}, based on $\sim$ 100 stars
in common between the two spectral libraries, is provided in
Table~\ref{conv}.

\begin{figure}
\includegraphics[scale=0.8,clip=true]{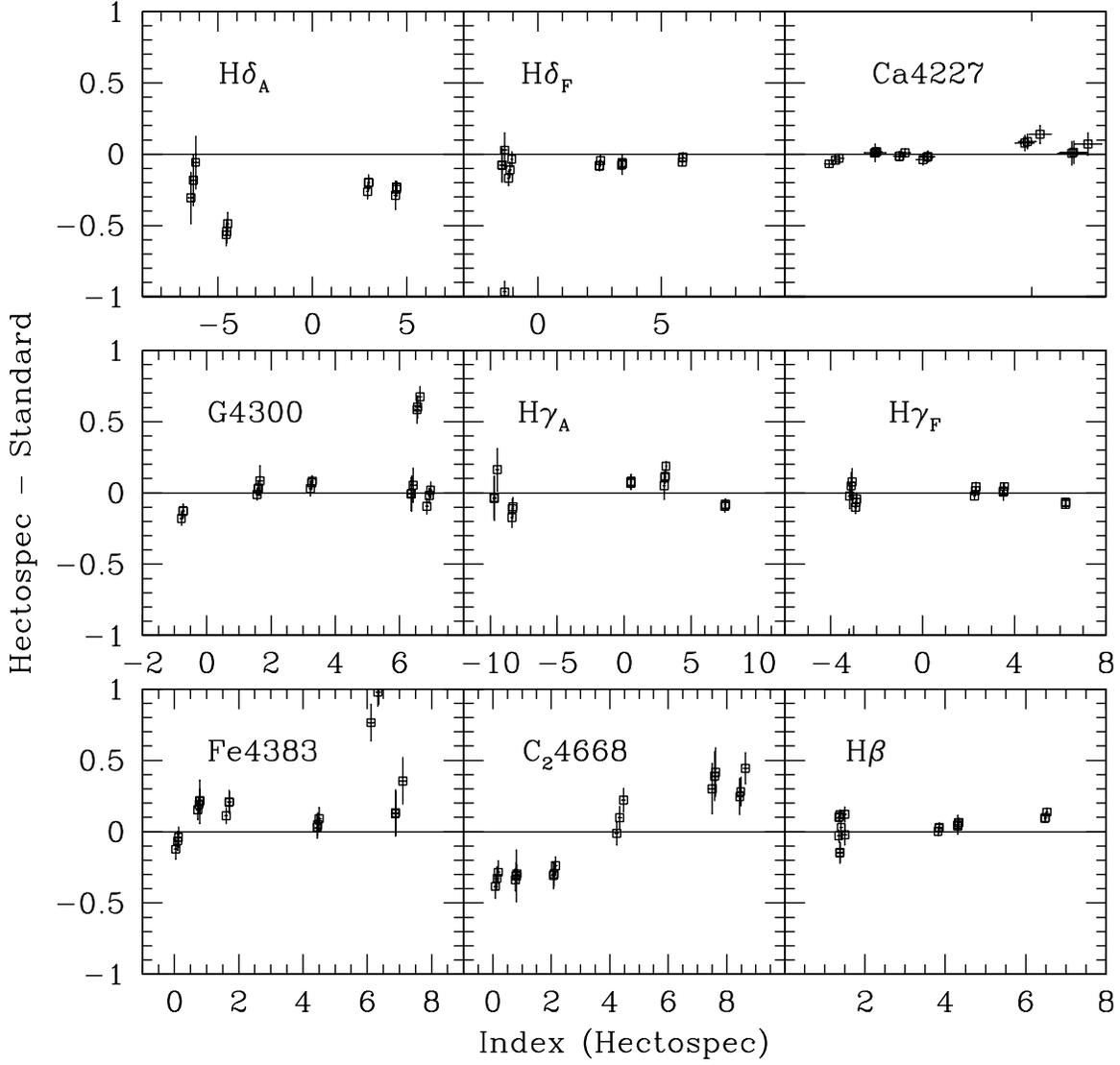}
\caption{Calibration of instrumental pseudo-equivalent widths into
the Lick system for the Hectospec integrated spectra of M~31 GCs,
using observations of Lick standards.  Zero points were
calculated by averaging the residuals for all indices, excluding
very strong deviants.  Standard Lick indices come from \cite{s07}.
Residuals are plotted as a function of instrumental index strength.
For all indices, the zero point shifts are relatively small, and
fairly well determined.
} \label{zm31a}
\end{figure}

\begin{figure}
\includegraphics[scale=0.8,clip=true]{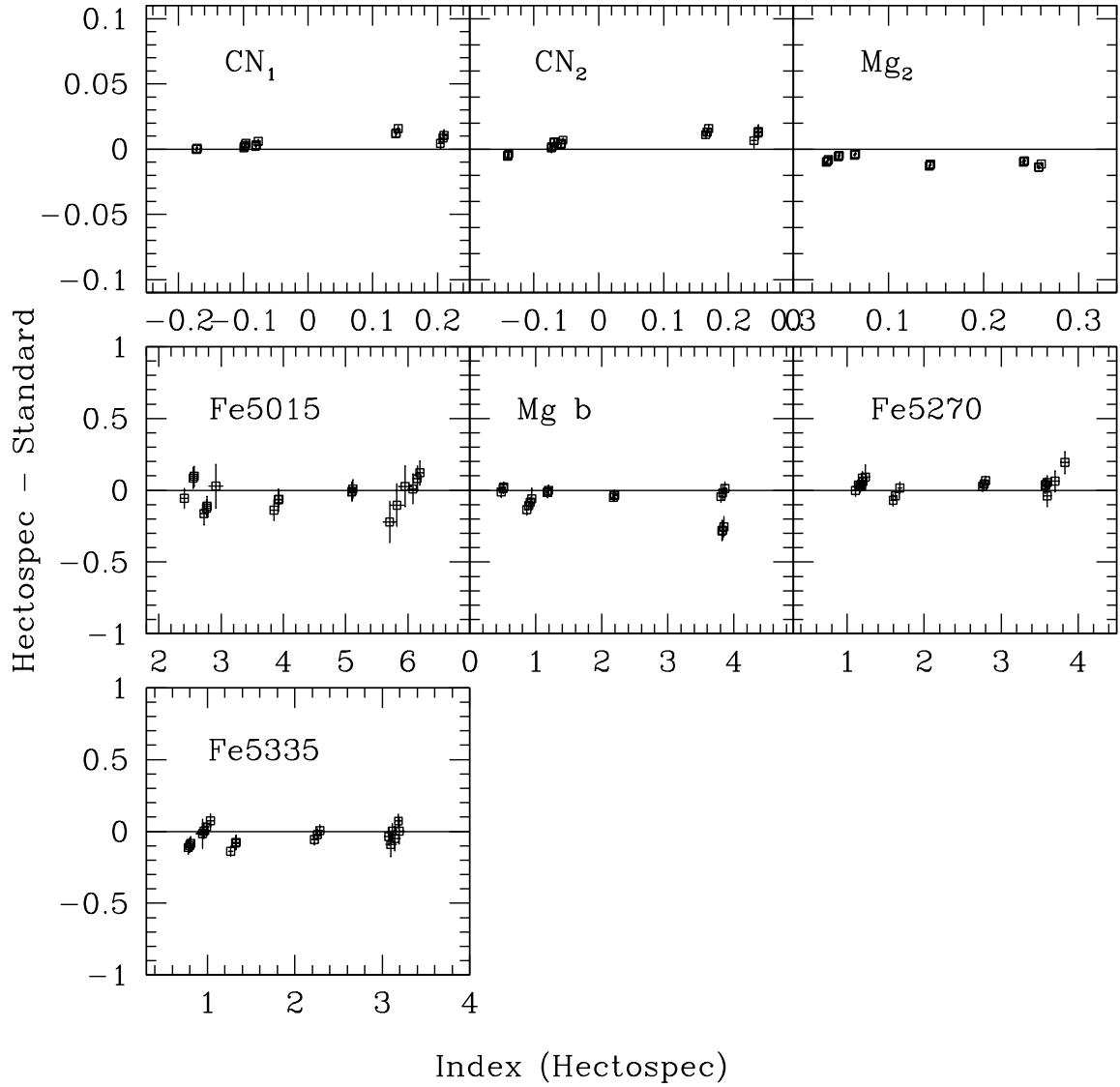}
\caption{Figure~\ref{zm31a} continued.  Note that, as in the case of the MW
GCs, (Figure~\ref{zerob}) the CN residuals are much smaller than the 0.05 mag
difference between M~31 and MW GCs, found by \cite{bu84}.}  \label{zm31b}
\end{figure}

\section{MW vs M~31 globular clusters in index-index space}
\label{analysis}

\begin{figure}
\includegraphics[angle=0,scale=0.8]{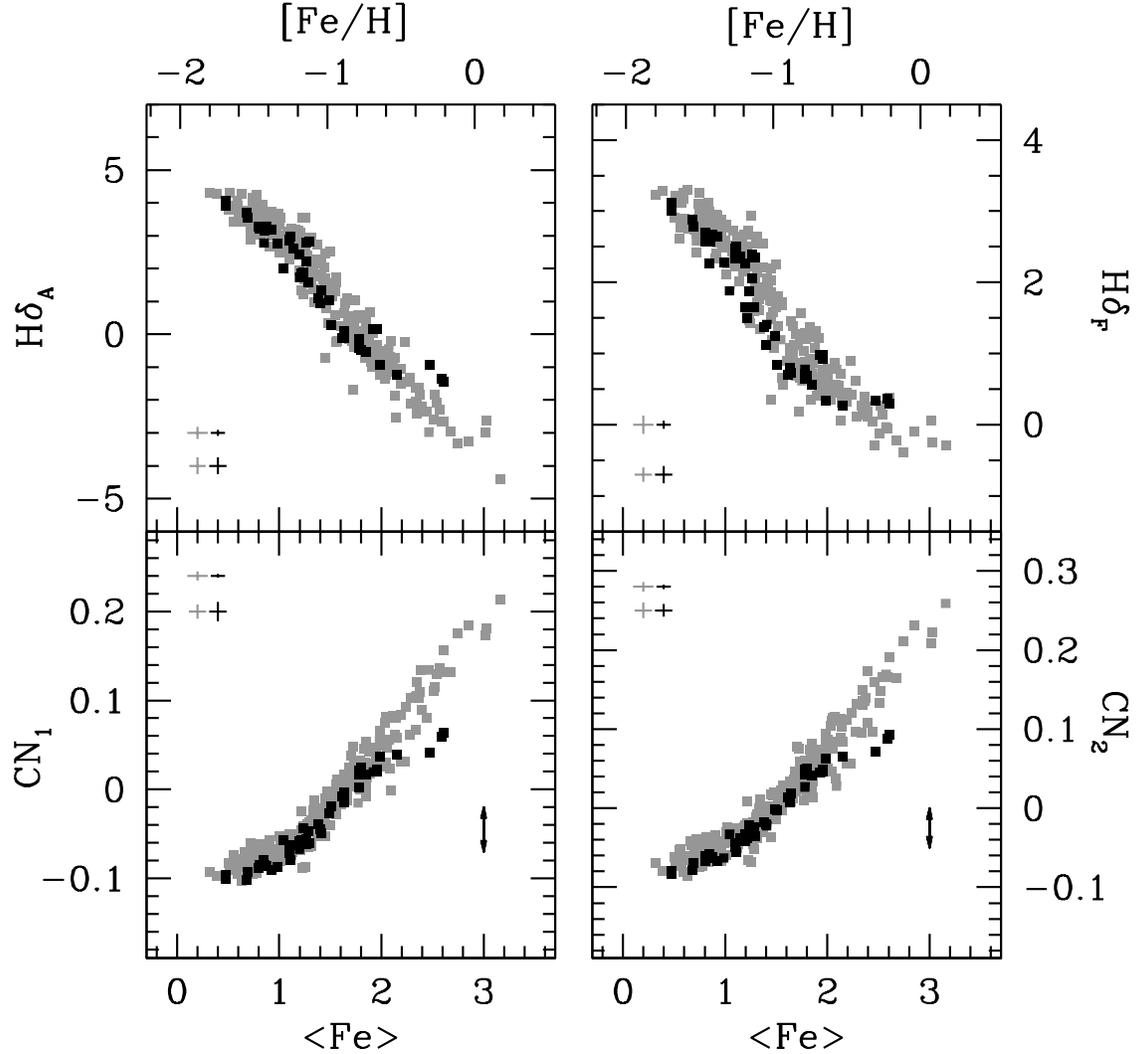}
\caption{ Lick indices for M~31 (gray) and MW (black) GCs.
Only GCs for which the error in $\langle$Fe$\rangle$ is smaller
than 0.15 ${\rm\AA}$ are plotted.  
Two sets of error bars are shown for
each GC system in the upper/lower left corner of each panel.
The upper set represents the average uncertainty in
individual index measurements, as estimated by {\tt lick\_ew}, using
the prescription from \cite{ca98}.  The lower set 
represents the zero point uncertainties determined as described in
Section~\ref{absind}.
The average $\langle$Fe$\rangle$ index is a
good indicator of [Fe/H], and a $\langle$Fe$\rangle$--[Fe/H] relation derived by
\cite{ca10} is used to establish the [Fe/H] scale at the top of the
upper panels.  That relation is strongly non-linear for [Fe/H]$\simless
-1.5$ and [Fe/H]$\simgreater 0.0$.  
Within the uncertainties, there
is no systematic difference between the two GC systems
in either $H\delta$ or CN indices---certainly nothing as high as
0.05 mag, as found by \cite{bu84}.  The arrows in the lower right
of the bottom panels indicate the size of the effect found by
\cite{bu84}.  The exceptions are the two most metal-rich MW GCs,
NGC~6528 and 6553 (where the former is represented by the two highest
$\langle$Fe$\rangle$ data points---see text).  These GCs are systematically
weaker in CN and stronger in H$\delta$, than M~31 GCs of the
same [Fe/H]---see discussion in text.  }

\label{inda}
\end{figure}

\begin{figure}
\includegraphics[angle=0,scale=0.8]{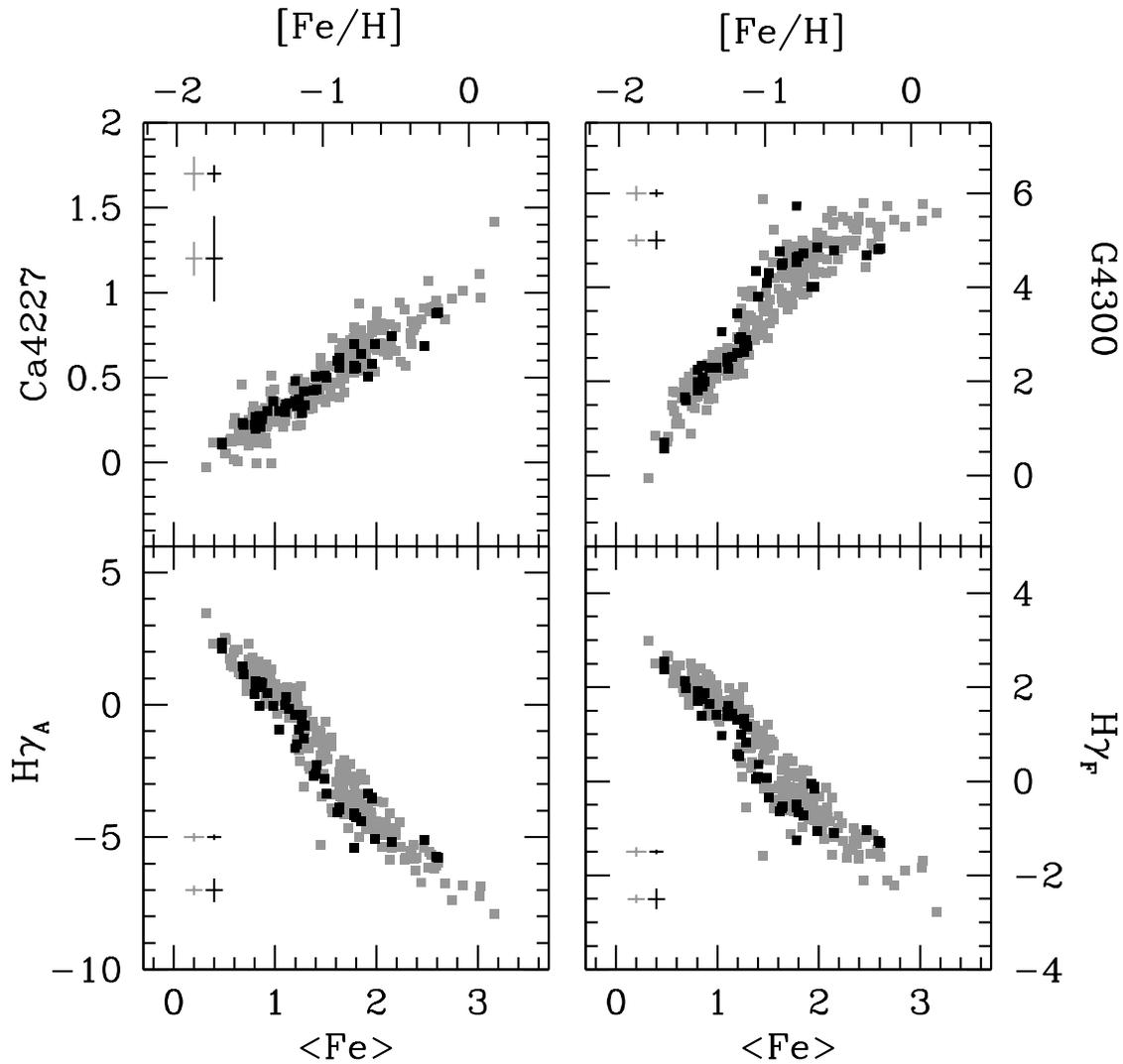}
\caption{Same as Figure~\ref{inda} for another set of line indices.  Again,
no significant differences between the two GC systems in index-index
space is found, again, with the exception of NGC~6528 and 6553, which are
too strong in $H\gamma$ and too weak in G4300, compared to M~31 GCs of
the same [Fe/H].  A few MW clusters seem to deviate from the trends
established by the remaining Galactic objects---see discussion in text.
}
\label{indb}
\end{figure}

\begin{figure}
\includegraphics[angle=0,scale=0.8]{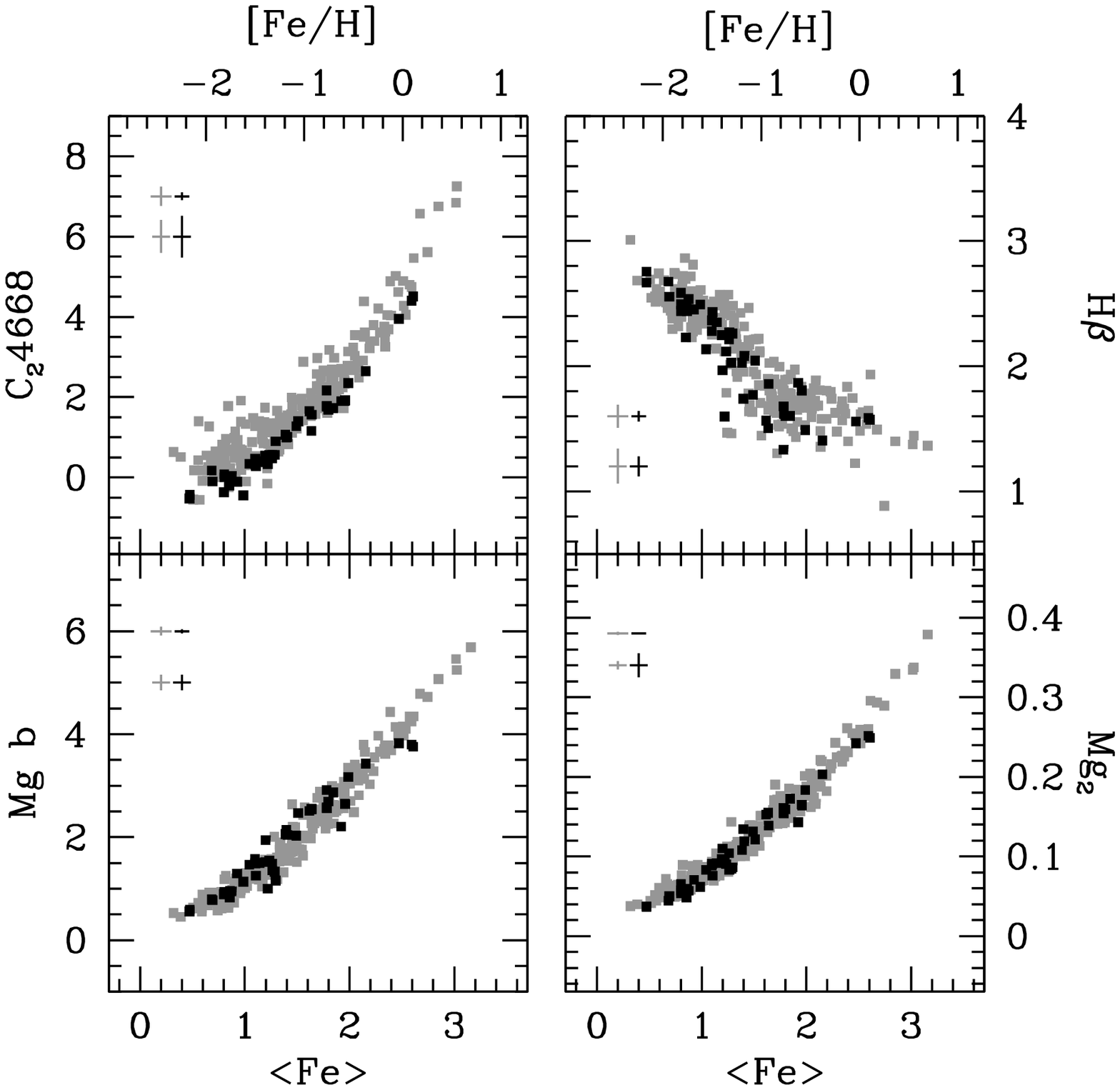}
\caption{Same as Figure~\ref{inda} for another set of line indices.  No
significant difference between the two GC systems is detected in $H\beta$ or
Mg indices.  There is a hint that $C_24668$ may be stronger in M~31 GCs
than in their MW counterparts of same [Fe/H], which would be indicative of
a higher carbon abundance in M~31 GCs.  However, the difference is not
larger than the zero-point uncertainty for this index, which is
particularly large for both data sets.  Unlike in the case of
the bluer Lick indices, the two metal-rich MW GCs do not differ from
their M~31 counterparts in any of the indices displayed here.  
}
\label{indc}
\end{figure}

The two systems of GCs are compared in index-index
space in Figures~\ref{inda}-\ref{indc}.  We plot all indices against
$\langle$Fe$\rangle$, which is the average of the two Lick indices Fe5270 and
Fe5335.  \cite{s07} showed that this combination of indices is a
good indicator of [Fe/H], being only weakly sensitive to variations
of other elemental abundances, and being only mildly sensitive to
age and horizontal branch (HB) morphology.  A relation between
$\langle$Fe$\rangle$ and [Fe/H] was derived by \cite{ca10}, on the basis of 31
Galactic GCs with [Fe/H] determinations from abundance analysis of
high resolution spectra of individual GC member stars by \cite{ca09},
\cite{ki03}, and \cite{cg97}.  Iron abundances in the two latter
papers were converted to the \cite{ca09} scale using the transformations
provided in their Appendix.  That relation is used to plot an
approximate [Fe/H] scale on the top of the upper panels in
Figures~\ref{inda}-\ref{indc}.  Due to non-linearities in the
$\langle$Fe$\rangle$--[Fe/H] relation, the [Fe/H] scale in those plots is only
approximate, but is useful to guide the eye.  For clarity and
accuracy, M~31 GCs for which the error in $\langle$Fe$\rangle$ is greater than
0.15 ${\rm\AA}$ are excluded from the plots.  This removes all young
clusters from the M~31 sample, because the S/N of the spectra for
these lower-mass, fainter clusters, is indeed on average lower than
that of their older, more massive counterparts.  Most importantly,
the absence of young and intermediate-age GCs in the M~31
sample means that age effects are, to first order, negligible in
metal-index vs. metal-index plots in Figures~\ref{inda}-\ref{indc}---though
they may be detectable in plots involving Balmer lines.  In this
way, we can perform a reliable model-free assessment of differences
(or lack thereof) between the old M~31 and MW GCs in metallicity
and abundance-ratio spaces.  Two sets of error bars are shown for
each GC system in the upper/lower left corner of each panel.
The upper set of error bars represents the average uncertainty in
individual index measurements, as estimated by {\tt lick\_ew}, using
the prescription from \cite{ca98}.  The lower set of error bars
represents the zero point uncertainties determined as described in
Section~\ref{absind}.

Before moving to the main focus of this paper, which is a comparison
of the strengths of CN indices in M~31 and MW GCs, we briefly mention
a few interesting features of Figures~\ref{inda}-\ref{indc}.  The
first obvious feature is the systematic difference between the two
GC samples in C$_24668$, where M~31 GCs are systematically stronger
by $\sim$ 0.5 ${\rm\AA}$.  We note that the calibration of this
index into the Lick system is relatively uncertain, as indicated
by the error bars on the top left panel of Figure~\ref{indc}, whose
size is in fact comparable to the shift between the two samples.
If taken at face value, this discrepancy could indicate a higher
[C/Fe] in M~31 GCs than in their MW counterparts.  However, the
data for the other carbon-sensitive index in this study, G4300, are
not consistent with that interpretation (Figure~\ref{indb}).  For
[Fe/H] $\simless$ --1.2, the two samples seem to have identical
G4300 values, and for higher [Fe/H] MW GCs actually have stronger
G4300 than their M~31 counterparts.  The latter is most likely not
associated with a difference in [C/Fe], but rather with differences
in age indicators between metal-rich MW and M~31 GCs, which are not
completely understood (see discussion in Section~\ref{thecase}).
We therefore conclude that the differences seen in C$_24668$ are
probably a spurious effect due to uncertainties in the calibration
of that index.  This topic will be revisited in a forthcoming paper,
where abundance ratios for both samples will be determined and
analyzed.

We finally point out that systematic differences between M~31 and
MW GCs with [Fe/H] $\simgreater$ --1.2 can be seen in
Figures~\ref{inda} and \ref{indb} in the strengths of their $H\delta$
and $H\gamma$ indices.  Interestingly, no such systematics are seen in
$H\beta$.  A comparative analysis of Balmer line strengths in the two
GC samples is the topic of another forthcoming paper.


\subsection{Fe vs CN indices} \label{fecn}

We focus on the bottom panels of Figure~\ref{inda}, which compare
the two GC samples in $\langle$Fe$\rangle$-CN space.  Note that the two CN
indices displayed in this Figure behave in exactly the same way.
This is because these indices measure the same spectral feature,
which is a combination of $\Delta v$ = 0 and --1 vibrational bands
of the CN violet electronic transition ($B_2\Sigma-X^2\Sigma$),
spreading roughly between 4100 and 4220 ${\rm\AA}$ \citep{wa62}.
The only difference between the two indices is in the blue
pseudocontinuum, which in the case of the $CN_2$ index is slightly
narrower.  Because our results are independent of which of the two
indices is used, we throughout this paper sometimes refer to {\it
CN indices} as both $CN_1$ and $CN_2$ Lick indices.

In this section, we concentrate our attention on GCs more
metal-poor than [Fe/H] $\sim$ --0.4, leaving the discussion of the
more metal-rich GCs for Section~\ref{thecase}.  The most obvious
feature in these plots is the slight difference between the two
GC samples in CN strength.  We find an average CN difference
between M~31 and MW GCs of about 0.01 mag, which is the same
size as the zero-point uncertainties in both sets of measurements.
This difference is also 1/5 of what was found by \cite{bu84}
(indicated by arrows in the lower right of the bottom panels).  In
other words, spectra of MW and M~31 GCs of the same $\langle$Fe$\rangle$
(same [Fe/H]) seem to have essentially the same CN strength (we
discuss the deviant, high-metallicity, MW GCs below).  This is in
disagreement with the results of \cite{bu84}, \cite{be04}, and
\cite{pu05}, who measured the same indices in smaller samples of
lower S/N spectra, finding evidence for a difference between the
two GC systems, in that M~31 GCs have stronger CN than their
MW counterparts by about 0.05 mag.  Similar CN differences between
MW and M~31 GCs were also suggested by \cite{bh90}, who measured
the strengths of the same CN bands studied in this paper, in a
sample consisting of tens of M~31 and MW GCs.  Similar differences
were also found by \cite{da90}, who measured the strength of CN
bands in NIR spectra of four M~31 GCs.  Because no difference was
found between the two GC systems in a carbon-sensitive index (the
G band), the above authors suggested that the CN differences were
probably due to M~31 GCs having higher nitrogen abundances than
their MW counterparts.  Just how much higher should the [N/Fe]
values be in M~31 than in MW GCs was not known, as such an
estimate required models that were then unavailable.  Nevertheless,
that suggestion was later bolstered by measurements of the near-UV
NH band at 3360${\rm\AA}$ by \cite{po98}, \cite{lb03}, and \cite{bu04}.
In particular, \cite{bu04} found NH3360 to be stronger in M~31 than
in MW GCs by over $\sim$ 2 ${\rm\AA}$ (see discussion in
Section~\ref{nhuv}).

\begin{figure}
\includegraphics[angle=0,scale=0.8]{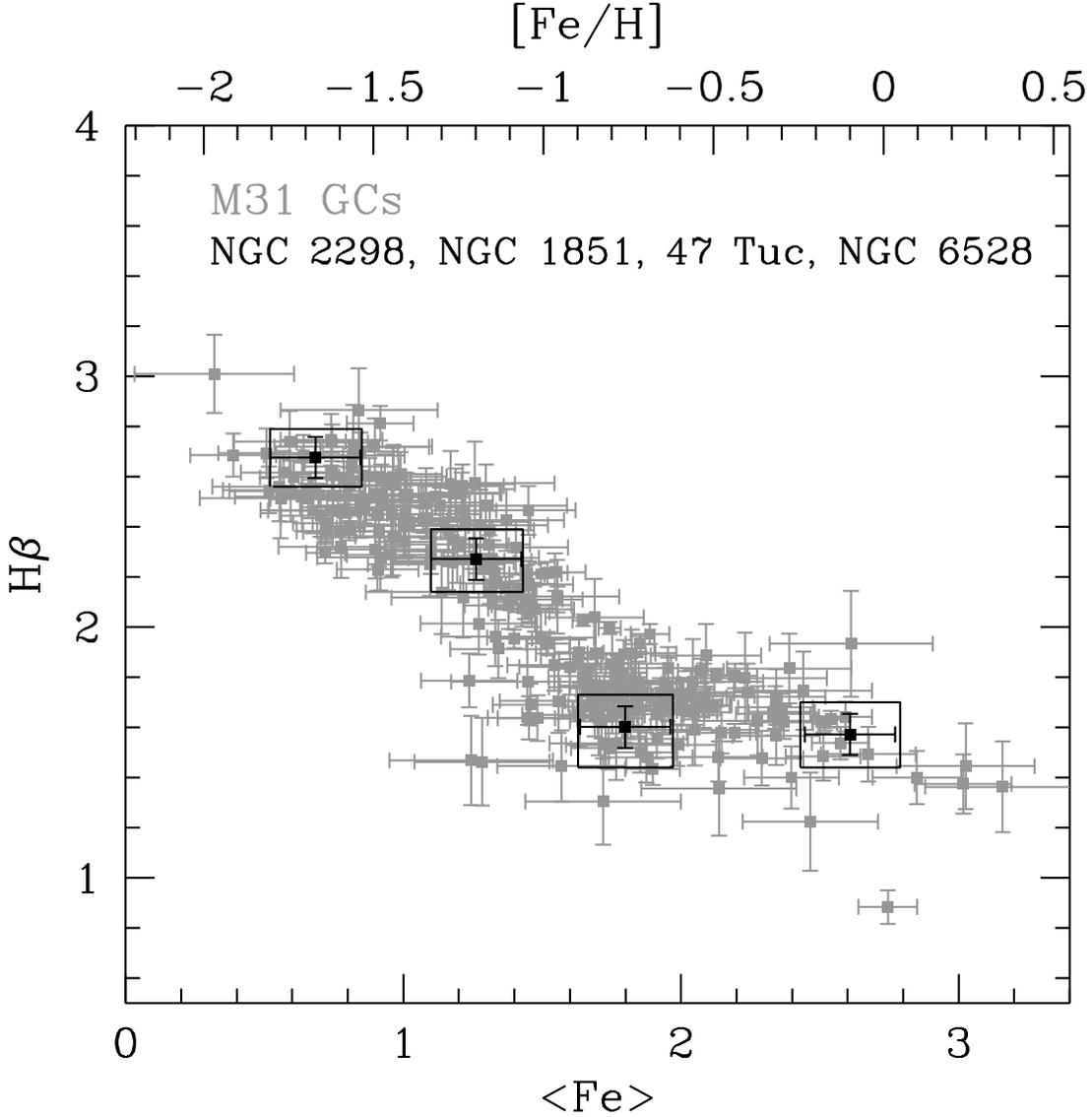}
\caption{This figure illustrates our criteria to select MW and M~31
GCs for spectral comparison.  Data in gray represent measurements
in M~31 GC spectra.  The black data points represent the measurements
and error bars for four MW GCs, spanning a wide range of [Fe/H].
The boxes surrounding the MW GC data points represent the index  bounds
for the search of similar GCs in M~31.  Those M~31 GCs that fall
inside each search box are deemed similar to their MW counterparts in terms
of [Fe/H] and a combination of age and HB morphology.
The spectra of these M~31 GCs are stacked and compared with
their MW counterparts, as shown in subsequent Figures.
}
\label{lookalike}
\end{figure}

\begin{figure}
\includegraphics[angle=0,scale=0.8]{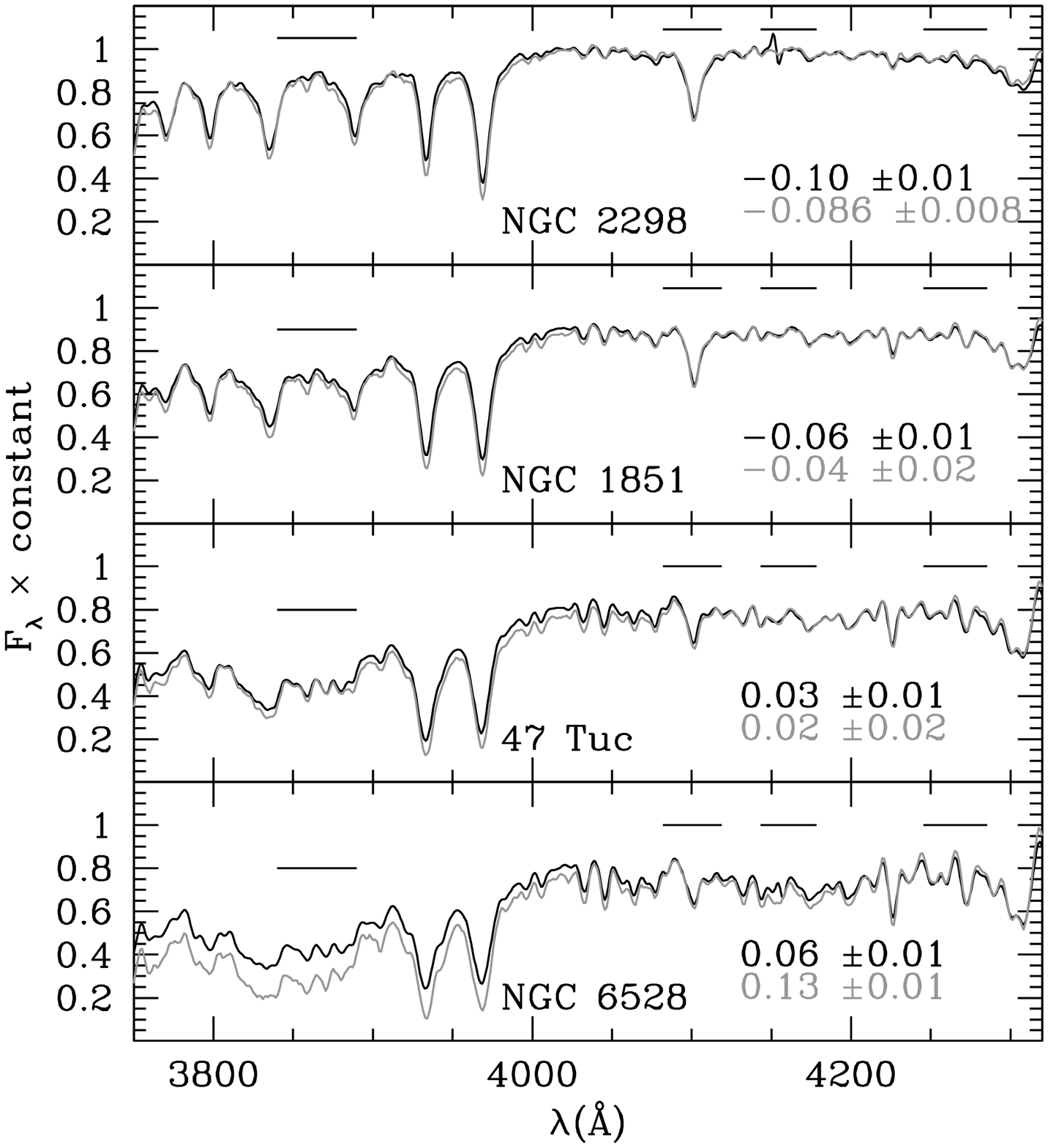}
\caption{Comparison of MW and M~31 GCs spectra selected according
to the criteria described in Figure~\ref{lookalike}.  MW GCs
are plotted in black, M~31 GCs in gray.  Each spectrum was multiplied by a
constant, so that the average flux within the red pseudocontinuum of the
Lick CN index are approximately the same.  MW GC spectra
were smoothed in order to match the lower resolution of M~31 spectra.
$CN_1$ indices measured in the MW spectra and the average index
values for M~31 GCs are displayed at the lower right corner
of each panel.  Horizontal bars at upper right of each panel represent
the pseudo-continuum windows and passband of the $CN_1$ index.  The
horizontal bar towards shorter wavelengths indicates the approximate
span of the CN violet bands.  Except for NGC~6528, there is no
indication for a difference in CN-band strength between MW and M~31
GCs.  See discussion in the text.
}
\label{spectra}
\end{figure}

Our result is then rather surprising, in view of the relatively
large body of counter-evidence summarized above, which led to the
(far-reaching) conclusion that M~31 and MW GCs have different
nitrogen abundances.  Considering the significance of the zero point
uncertainties discussed in Section~\ref{absind}, we decided to take
advantage of the high S/N of our data to inspect visually the spectra
of M~31 and MW GCs.  That required selecting GCs with the same
[Fe/H] and same combination of age and HB morphology.  We proceeded
as follows.  We first selected four MW GCs spanning a wide range
of metallicities for the comparison: NGC~2298, NGC~1851, 47~Tuc,
and NGC~6528, with [Fe/H] = -1.96, -1,18, -0.76, and +0.07 \citep{ca09}.
Next, we looked for M~31 GCs with similar parameters.  The search
is illustrated in Figure~\ref{lookalike}, where data for those four
GCs are overplotted on top of M~31 GC measurements in the
$\langle$Fe$\rangle \times\, H\beta$ plane.  \cite{s04b} and
\cite{s07} showed that these indices are chiefly sensitive to [Fe/H],
age, and HB morphology.  Therefore, selecting M~31 GCs with similar
$\langle$Fe$\rangle$ and $H\beta$, one should obtain a sample with
[Fe/H] and age/HB-morphology that is similar to those of the four
MW GCs of choice.  The boxes surrounding the latter four data points
illustrate our selection of four sub-samples of GCs which are the
counterparts of NGC~2298, NGC~1851, 47~Tuc, and NGC~6528 in M~31,
according to that (limited) set of stellar population parameters.
Clusters with low S/N spectra or unusual-looking spectral shapes
(indicative of uncertain flux-calibration) were removed from the
sub-samples.  The IDs of the GCs selected for each metallicity bin
are listed in Table~\ref{lookalikes}.  Intercomparison of these
spectra should reveal any CN differences that may be masked by
zero-point uncertainties in our Lick index measurements.  Very high
S/N stacked spectra of M~31 GCs within each of the four metallicity
groups were generated, which are compared with those of their MW
counterparts in Figure~\ref{spectra}.  The latter spectra were
smoothed to match the lower resolution of the M~31 GCs spectra.  In
all panels, the wavelengths of the pseudocontinuum and passband
windows defining the Lick $CN_1$ index are indicated as horizontal
bars, and $CN_1$ measurements taken on these spectra are shown in
the lower-right corner of each panel.  Another bar towards shorter
wavelengths indicates the approximate span of the (stronger) violet
CN bands, which go from roughly 3840 to 3890 ${\rm\AA}$ \citep{wa62}

We first focus on the spectra of NGC~2298, NGC~1851, and 47~Tuc,
leaving the discussion of NGC~6528 to Section~\ref{thecase}.  The
spectra of these three clusters are remarkably similar to those of
their M~31 counterparts, except for slight slope differences, which
can be ascribed to uncertainties in flux calibration and reddening.
There are also small differences in the strengths of Ca HK and
higher order Balmer lines, which may be due to a combination of
sky-subtraction uncertainties and genuine differences in the relative
numbers of the hottest stars and their temperature distributions
(a topic that is further discussed in a forthcoming paper).  However,
in confirmation of our conclusions from Figure~\ref{inda}, there
are {\it no differences} in CN strengths between these two sets of
spectra, either in the 4170 $CN_1$ band, or in the violet CN band.
The latter is particularly important because the violet CN band is
more sensitive than its blue counterpart to variations of CN
concentration in stellar atmospheres, further reinforcing the notion
that there are no differences in CN strengths between M~31 and MW
GCs.  For NGC~2298, NGC~1851, and their M~31 counterparts, the
spectral shape and the strengths of the lines included in the
bandpass of the $CN_1$ index suggest that blue CN lines are indeed
very weak, if not altogether absent, from the spectrum.  Therefore,
we suggest that any differences between M~31 and MW spectra in this
low metallicity regime cannot be ascribed to differences in CN
strength (the 0.014 mag difference measured in $CN_1$ between
NGC~2298 and its M~31 counterparts is mostly due to the blip in the
spectrum of NGC~2298, which is associated with a bad CCD column,
as discussed in Section~\ref{absind}).  These comparisons also
suggest that zero-point uncertainties in $CN_1$ measurements are
negligible.

Having found no compelling reasons in our own data to conclude that
there is any significant CN-strength difference between M~31 and
MW GCs, we examine other data sets published in the literature,
in an attempt to understand the contrast between our results and
those from other studies.  This is the topic of the next three
sub-sections.

\subsubsection{The Puzia et al. data} \label{pudata}

The result above is further confirmed by inspection of other data
sets, such as those published by T. Puzia and collaborators.
\cite{pu02} obtained integrated spectra for Galactic GCs, with a
resolution of 6.7 ${\rm\AA}$, using the Boller \& Chivens spectrograph
attached to the 1.52~m telescope at ESO/La Silla.  The observational
strategy followed by \cite{pu02} differs from that of \cite{s05}
in two important aspects.  The first concerns the sampling of the
extended GC spatial distributions.  While \cite{s05} drift-scanned
the target GCs within their core radii, \cite{pu02} chose instead
to take a number of exposures with the slit placed at different
positions on the GCs.  The second regards sky subtraction.  \cite{s05}
scanned regions of sky around the GCs generating 1D background
spectra which they subtracted from the GC extracted 1D spectra.
\cite{pu02} followed two different procedures.  On one hand, like
\cite{s05}, they subtracted 1D background spectra from their extracted
cluster spectra (background ``modeling'').  On the other they used
narrow windows at the slit ends to sample the sky background,
subtracting it in the process of extracting the 1D spectrum (background
``extraction'').  They argue that background ``modeling'' yields
unreliable results, because it is difficult to model the light
around the clusters, mostly due to superposition of different stellar
populations, and reddening variations at 1.5$\arcmin$ scales.  The
\cite{pu05} M~31 GC data were obtained by \cite{pe02} using the
Wide Field Fibre Optic Spectrograph on the 4 m William Herschel
Telescope at La Palma, Canary Islands.  Those data resemble our own
M~31 data in terms of spatial sampling, although their resolution
and S/N is somewhat lower.

\begin{figure}
\includegraphics[angle=0,scale=0.8]{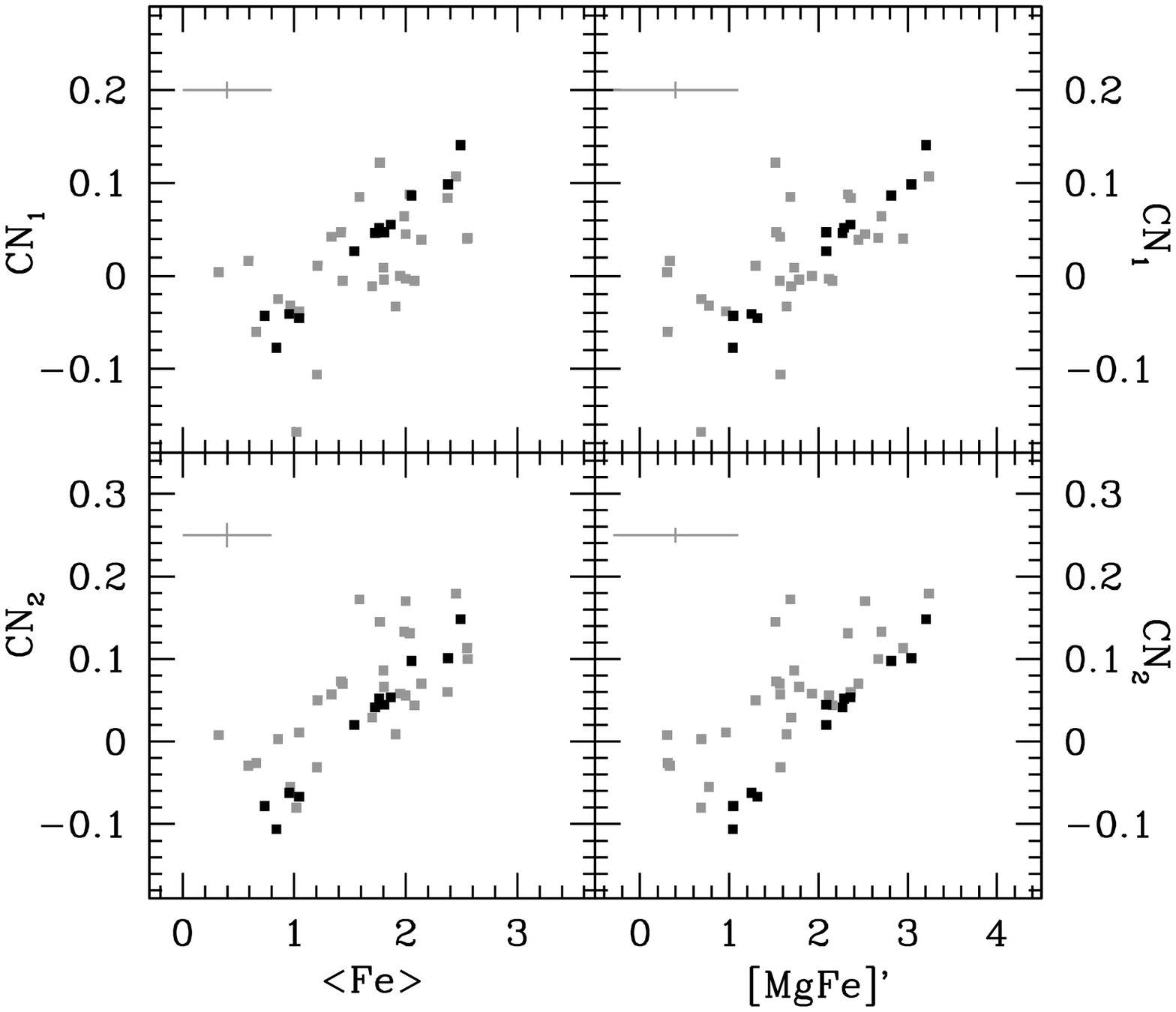}
\caption{Data from \cite{pu05} for M~31 GCs are compared with data
from \cite{pu02} for MW GCs.  Taking error bars at face value,
there is formally no difference between CN strengths in M~31 and
MW GCs according to these data sets.  The lower-right panel suggests a
possible difference, with the M~31 GCs being stronger than MW GCs
by an amount comparable to the 0.05 mag suggested by \cite{bu84}.  We
suggest that this is because of a difference in Mg $b$, rather than CN, as
argued in Figure~\ref{puziamgfe}.
}
\label{puziadata}
\end{figure}

In Figure~\ref{puziadata}, we inter-compare data from \cite{pu02}
and \cite{pu05} for MW and M~31 GCs, respectively.  Index measurements
are those in the \cite{wo94} system.  The mean index errors for
M~31 GCs, taken from \cite{pu05}, are represented by error bars in
each panel.  The nominal measurement errors from \cite{pu02} for
MW GCs are comparable to symbol sizes, and are not shown.  The left
panels show the Lick $CN_1$ and $CN_2$ indices plotted against
$\langle$Fe$\rangle$.  On both panels, M~31 and MW GCs occupy approximately the
same locus.  There is no compelling evidence for M~31 GCs
having stronger CN indices at fixed $\langle$Fe$\rangle$ (fixed [Fe/H]).  The
right panels show CN indices plotted against the [MgFe]' index,
which is defined as

\begin{equation}
[MgFe]' = \sqrt{Mg b (0.72 Fe5270 + 0.28 Fe5335)}.
\end{equation}

\begin{figure}
\includegraphics[angle=0,scale=0.8]{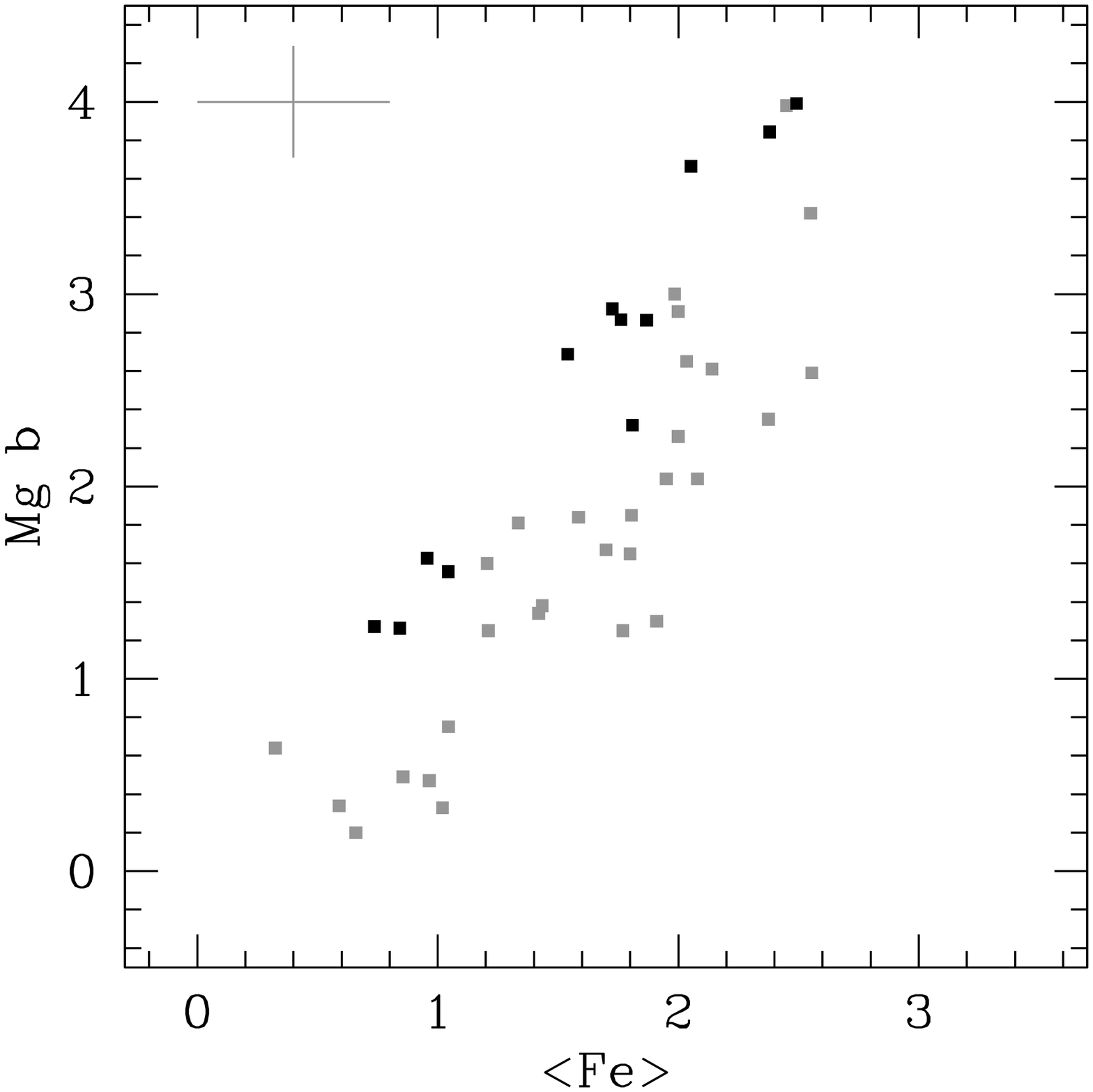}
\caption{Data from \cite{pu05} for M~31 GCs are compared with data from
\cite{pu02} for MW GCs.  The latter have stronger Mg $b$ for given
$\langle$Fe$\rangle$, suggesting that the differences between the two samples in the
lower right panel of Figure~\ref{puziadata} are in fact due to a
discrepancy in Mg $b$, not in CN.  We note that our data do not show any
differences between M~31 and MW GCs in Mg indices, and ascribe this
discrepancy in the data by Puzia and collaborators to possible calibration
uncertainties.
}
\label{puziamgfe}
\end{figure}

\noindent Again, no formal difference is found between the two GC
systems in these diagrams.  The MW GCs seem to occupy the lower
envelope of the distribution of M~31 GCs in the [MgFe]' vs. $CN_2$
plane.  However, since there are no differences in the $\langle$Fe$\rangle$
vs.  $CN_2$ plane, this mismatch is likely to be due to Mg $b$ differences
between the two GC systems in the data by Puzia and collaborators,
and {\it not} in their CN strengths.  The latter suspicion is
confirmed by Figure~\ref{puziamgfe}, where Mg $b$ and $\langle$Fe$\rangle$
from \cite{pu02} and \cite{pu05} are compared.  The MW data are
systematically stronger in Mg $b$ for fixed $\langle$Fe$\rangle$
than M~31 data, by about 0.5 ${\rm\AA}$.  Following a suggestion
by the anonymous referee, we generated plots similar to
Figure~\ref{puziamgfe}, replacing $\langle$Fe$\rangle$ by other Fe
indices, such as Fe4383, Fe4531, Fe5406, and Fe5709, and Mg $b$ by
Mg$_2$, and the same result is obtained: MW clusters are systematically
stronger than their M~31 counterparts in Mg indices at fixed Fe,
which suggests that the difference is due to Mg $b$ being stronger,
and not $\langle$Fe$\rangle$ being weaker, in MW clusters.  We note
that no such difference is found in our data between MW and M~31
(Figure~\ref{indc}), so we do not understand the origin of this
discrepancy in the data by Puzia and collaborators.  Comparing Mg
$b$ for MW clusters in common between \cite{pu02} and \cite{s07},
we find that indeed the former are on average stronger by $\sim
0.13~{\rm\AA}$.  On the other hand, comparing the \cite{pu05} data
for M~31 clusters with our own data for clusters in common, we find
their Mg $b$ values to be on average $\sim 0.3~{\rm\AA}$ weaker and
their $\langle$Fe$\rangle$ values to be $\sim 0.1~{\rm\AA}$ weaker.
Incidentally, we also found the $CN_1$ and $CN_2$ values in \cite{pu05}
M~31 GC data to be systematically stronger than our own values by
$\sim 0.02-0.03$ mag, which may also be partly responsible for their
results.  We note that the \cite{pe02} and \cite{pu05} M~31 GC data
do not seem to have been flux calibrated, and moreover their line
indices were not transformed to the Lick system through the comparison
of observations of Lick standards with tabulated values.  Therefore,
we have reasons to suspect that the \cite{pu05} index values may
be subject to systematic effects, which may explain
the discrepancies found in Figure~\ref{puziamgfe}.

In summary, we conclude that the data by \cite{pu02} and \cite{pu05}
are probably consistent with no difference in CN strength between
the M~31 and MW GCs.

\subsubsection{The \cite{bu04} data}  \label{nhuv}

\begin{figure}
\includegraphics[angle=0,scale=0.8]{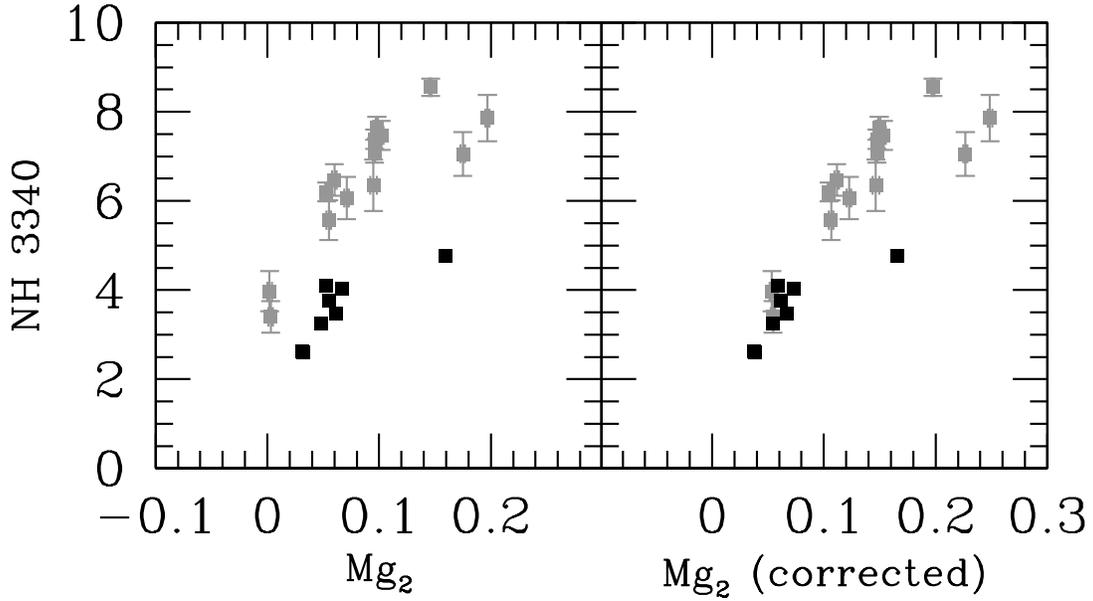}
\caption{{\it Left Panel:} Data from \cite{bu04} for M~31 (gray),
and MW GCs (black).  The data suggest that M~31 GCs have
stronger NH bands than their MW counterparts with same Mg$_2$.  {\it
Right Panel:} Same data, but with Mg$_2$ indices zero-point corrected,
to bring them to the same system as the data presented in this
paper, using GCs in common with the two samples.  Except for
the most metal-rich GCs in the MW sample (M~71), MW GCs 
seem to have NH strengths that are consistent with those of their
M~31 counterparts.  This plot suggests that there may be problems
with the zero-points of Mg$_2$ indices in \cite{bu04} data, and
when this effect is removed, they are in fact consistent with similar
NH strengths in MW and M~31 GCs.
}
\label{burstdata}
\end{figure}

Close inspection of other data sets also confirm our results.  For
instance, \cite{bu04} claimed that the UV NH band at $\sim$ 3360
${\rm\AA}$ is substantially stronger in the spectra of M~31 GCs
than in their MW counterparts.  \cite{bu04} argue that this is
further evidence for a difference in nitrogen abundance between
M~31 and MW GCs---the claimed physical origin for the alleged
CN differences between the two GC systems.  Figure~\ref{burstdata}
shows the data from \cite{bu04} for the NH 3360 and Mg$_2$ indices.
The left panel is a reproduction of the top panel of their Figure
7, which suggests that the NH 3360 band is much stronger in M~31
GCs than in their MW counterparts.  The mean indices for the
bulk of the MW sample in \cite{bu04} are Mg$_2$ $\sim$ 0.06 mag,
and NH 3360 $\sim$ 3.5 ${\rm\AA}$ (excluding M~71, the GC with
highest metallicity in that sample).  Clusters in M~31 with comparable
Mg$_2$ strength have NH 3360 $\sim$ 6 ${\rm\AA}$, which is almost
a factor of 2 stronger than their MW counterparts.

We compared the Mg$_2$ measurements in \cite{bu04} with our numbers
for GCs in common in the two studies.  From 14 M~31 GCs
in common between this work and \cite{bu04}, we determine a mean Mg$_2$
difference of 

\begin{equation}
Mg_2 (This\,\, work) - Mg_2 (B04) = 0.052 \pm 0.005\,\, mag
\end{equation}

The anonymous referee pointed out to us that a systematic difference
exists between the \cite{bu04} Mg$_2$ data for M~31 GCs and those
by \cite{tr98}, which is consistent with these numbers.

For the MW sample, there are unfortunately only 3 GCs in common
between the two studies: NGC~5904, 7078, and 7089, so zero point
differences cannot be determined as robustly.  Nevertheless, the
differences between Mg$_2$ measurements from \cite{bu04} and this
work, in the same sense as above, for these three GCs, are: 0.0089,
0.0062, and 0.0038 mag, respectively.  These are an order of magnitude
smaller than the zero-point differences for the M~31 sample, and
in fact are comparable to our zero point uncertainty from
equation~\ref{zpt}.  We apply these zero point corrections to the
\cite{bu04} Mg$_2$ data, and redisplay them on the right panel of
Figure~\ref{burstdata}.  One can see from this Figure that most of
the difference between the two samples disappears once the \cite{bu04}
Mg$_2$ indices are brought into consistency with our values.  This
analysis therefore suggests that most of the apparent discrepancy
between NH 3360 measurements in M~31 and MW GCs is in fact due to
uncertainties in the calibration of Mg$_2$ measurements to a common
system, and {\it not} to a real difference in NH strengths.

\begin{figure}
\includegraphics[angle=0,scale=0.8]{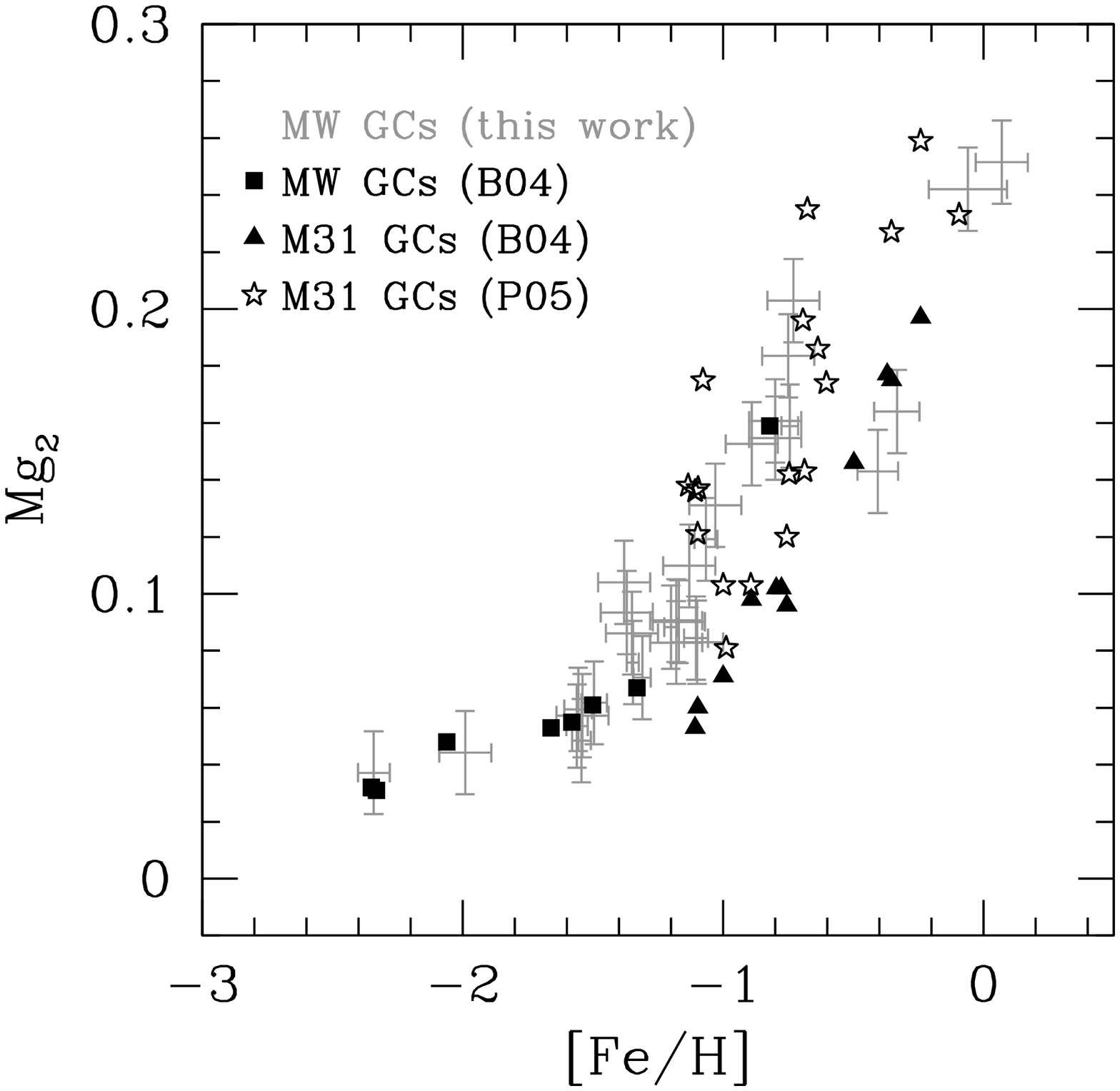}
\caption{Run of Mg$_2$ with [Fe/H] for MW and M~31 GCs.  Mg$_2$
comes from different sources: {\it Gray error bars:} MW GCs
from \cite{s05}; {\it solid squares:} MW GCs from \cite{bu04};
{\it solid triangles:} \cite{bu04} data for M~31 GCs; {\it
open stars} \cite{pu05} data for M~31 GCs.  Iron abundances
for M~31 GCs were determined using the \cite{s07} models, by
\cite{ca10}, and do not rely on Mg$_2$ measurements.  For MW
GCs, [Fe/H] comes from \cite{ca09}.  All the data for both
M~31 and MW GCs occupy the same locus in this plane, except for
the \cite{bu04} data for M~31 clusters, suggesting that there indeed
is a zero-point offset for this index in \cite{bu04} data.
}
\label{mg2feh}
\end{figure}

That there might be issues with the calibration of the Mg$_2$ index
into the Lick system is not entirely implausible, since this index
is measured on a wide baseline, being thus susceptible to uncertainties
in flux calibration.  However, a $\sim$ 0.05 mag difference seems
too large to be explained by spectrophotometric errors alone.
\cite{bu04} indices for M~31 and MW GCs were measured on
flux-calibrated spectra obtained, respectively, with the Blue Channel
spectrograph, at MMT, and the Boller \& Chivens spectrograph on the
Cassegrain focus of the Bok telescope.  Instrumental magnitudes
were converted to the Lick system using observations of standard
stars, in the usual way.  We do not find any obvious way in which
the calibration of Mg$_2$ measurements in \cite{bu04} data may be
faulty.  By the same token, the calibration of our own measurements
for that index into the Lick system seems to be quite robust (upper
panel of Figure~\ref{zm31b}).  However, an issue indeed seems to
be present with the \cite{bu04} Mg$_2$ indices for M~31 GCs.
This is further illustrated by Figure~\ref{mg2feh}, where Mg$_2$
measurements from different sources for M~31 and MW GCs are
plotted against [Fe/H].  Iron abundances for the M~31 GCs were
obtained by \cite{ca10} as described in Section~\ref{analysis}.  For
MW GCs, [Fe/H] comes from \cite{ca09}, whereas Mg$_2$ comes
from different sources: MW GCs from this work (gray error
bars), MW GCs from \cite{bu04} (solid squares), M~31 GCs 
from \cite{bu04} (solid triangles), and M~31 GCs from \cite{pu05}
(open stars).  Clusters in M~31 and the MW are expected to occupy
the same locus in this diagram, provided they have similar [Mg/Fe],
which is the case, as demonstrated by \cite{co09}, \cite{s11}, and
Figure~\ref{indb}.  As can be seen, the data by \cite{pu05} are in
good agreement with our MW data, and so are the MW data by \cite{bu04}.
However, the M~31 data by \cite{bu04} clearly depart from the overall
trend, towards lower Mg$_2$, suggesting a zero point offset of about
0.05 mag in that index.

We conclude that the NH 3360 feature has similar strength in the
spectra of M~31 and MW GCs of the same metallicity.  We suggest
that the assertion that it may be stronger in the spectra of M~31
GCs derived from a zero point offset in Mg$_2$ measurements
in \cite{bu04}.  M~31 and MW GCs of different metallicities
were compared in the NH vs Mg$_2$ plane, which led to a perception
that NH features seemed (artificially) stronger in M~31 than in MW
GCs.  Therefore, we conclude that the data do not require M~31
GCs to have enhanced N abundances, relative to their MW
counterparts of same metallicity.  In retrospect, this is not
surprising.  Looking carefully at the original data by \cite{bu84}
and \cite{bh90}, one notices that the bulk of the difference between
the two samples happens in the low metallicity regime.  For instance,
looking at Figure 5l in \cite{bu84}, one finds that CN differences
between M~31 and MW GCs happen for GCs with Mg$_2 \,
\simless \, 0.1$, which, according to Figure~\ref{mg2feh}, corresponds
to [Fe/H] $\simless$ --1.0---the same actually applies to the
\cite{bh90} data.  Recall that we concluded, in our discussion of
GC blue spectra from Section~\ref{fecn} (Figure~\ref{spectra}),
that CN 4170 is hardly present in GCs with such low metallicities,
suggesting that the measured differences may be associated with
uncertainties in the calibration of the CN and/or Mg indices.

\subsubsection{The Beasley et al. data}

We also examine the claim by \cite{be04} that CN is enhanced in
M~31 GCs.  \cite{be04} obtained high S/N integrated spectra
of $\sim$ 30 M~31 GCs with the LRIS spectrograph
connected to the Keck I telescope.  The spectra were reduced following
standard procedures, and the authors report flux calibration
uncertainties of the order of 10\%.  Indices were measured in M~31
GC spectra and converted to the Lick system using observations of
standard stars.  These indices were compared with measurements taken
on Galactic spectra by \cite{co98} and \cite{pu02}.  

Figure 3 of \cite{be04} shows very clearly that there are no
differences in the Lick CN$_2$ index between M~31 and MW GCs of
same metallicity.  However, \cite{be04} claim that instead there
are differences in the CNB index, defined by \cite{bh86}, which
measures the strength of the violet CN 3883 ${\rm\AA}$ feature.
Figure 4 of \cite{be04} suggests that this index is stronger in
M~31 GCs than in their MW counterparts by as much as 0.1 mag.
Because this result is in disagreement with their own measurements
of the Lick CN indices, \cite{be04} provide a comparison between
two well observed spectra from the M~31 and the Galaxy: those of
255-280 and NGC~6441, respectively.  This comparison, in the form
of a ratio spectrum, is shown in their Figure 5, where residuals
are seen at the positions of the Lick CN$_2$ and CNB indices, as
well as the Ca II H and K lines.  The latter residual is particularly
strong.  However, closer inspection of that figure shows that the
``feature'' in the ratio spectrum below $\sim$ 4000 ${\rm\AA}$ is
very broad with FWHM $\sim$ 200 ${\rm\AA}$, whereas the CN violet
feature is certainly not broader than 100 ${\rm\AA}$ (see
Figure~\ref{spectra}).  This suggests that the difference may be
instead due to some systematic flux calibration effect, rather than
to a real difference in CN strength.

\begin{figure}
\includegraphics[angle=0,scale=0.8]{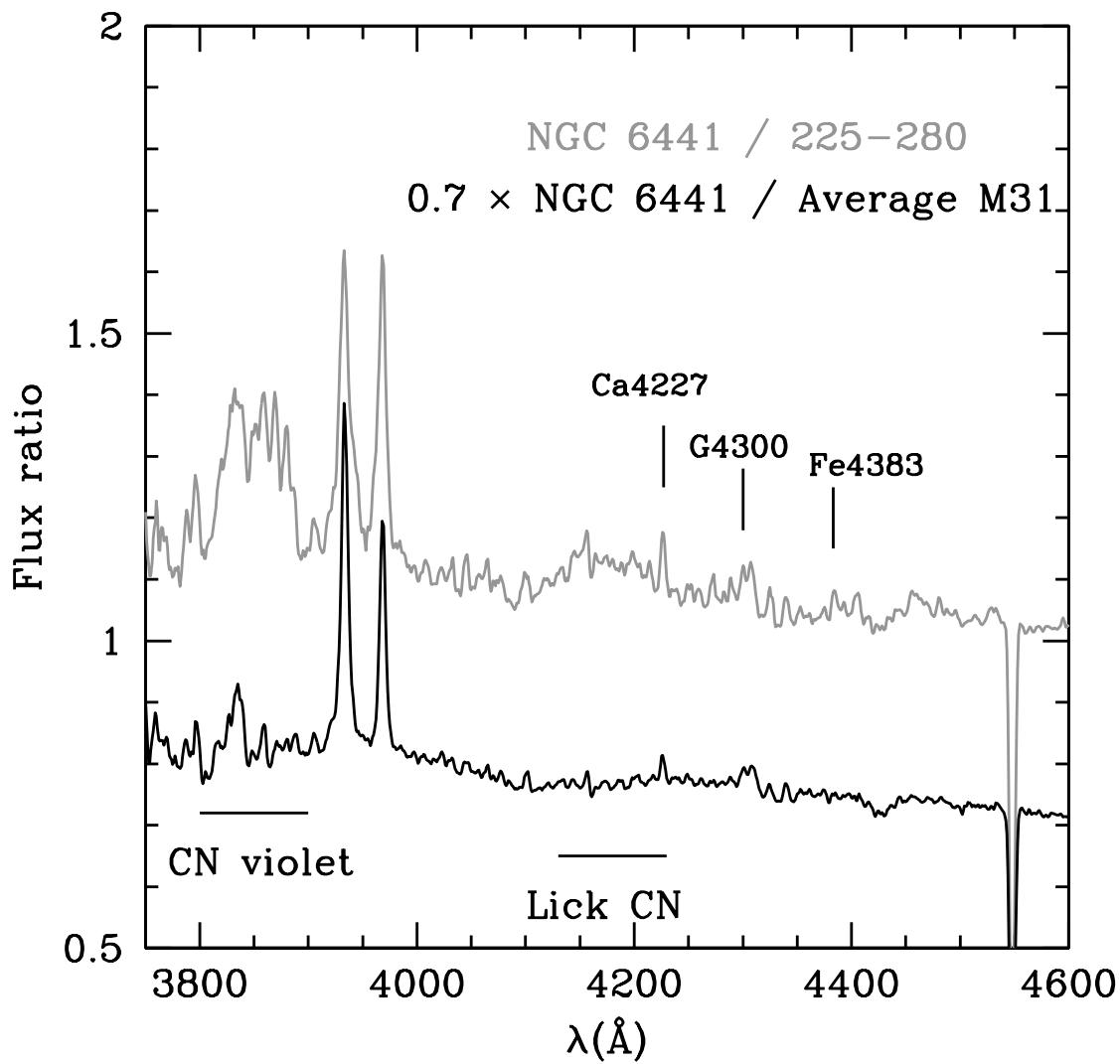}
\caption{Ratio between the spectra of NGC~6441 and M~31 GCs.  The gray
line shows the ratio between NGC~6441 and 225-280 \citep{be04} and the
black line that between NGC~6441 and the average spectrum of M~31 GCs
with same metallicity and age/HB-morphology.  The gray line shows that,
contrary to the suggestion by \cite{be04}, 225-280 is more metal-rich,
rather than CN-enhanced relative to NGC~6441.  Residuals due to CN bands
essentially vanish when NGC~6441 is compared with GCs of same
metallicity.  See text
}
\label{ratio}
\end{figure}

This is further verified in Figure~\ref{ratio}, where the same ratio
spectrum obtained with our own data for NGC~6441 and 225-280 (gray
lines) is compared with the ratio between the spectra of NGC~6441
and the average spectrum of M~31 GCs with similar [Fe/H] and
age/HB-morphology as NGC~6441 (black line).  Focusing first on the
gray line, we note that our version of the ratio between the spectra
of NGC~6441 and 225-280 is very different from that shown in Figure
5 of \cite{be04}.  The residual at the position of the CN violet
band has a completely different shape than that shown in Figure 5
of \cite{be04}.  While in our case the residual keeps a strong
resemblance to the shape of the CN violet band, in \cite{be04} it
is indeed much broader and stronger, suggesting that it is very
likely due to flux calibration and or sky-subtraction uncertainties.
Such uncertainties are not unusual at the blue end of the spectrum,
which is often characterized by low S/N, due to a combination of
low detector quantum efficiency, high atmospheric extinction and
sky background, and the overall redness of the GCs.  Nevertheless,
the fact remains that our ratio spectrum does show a strong residual
at the position of both the violet and the Lick CN bands.  This is
explained, however, by the fact that 225-280 is {\it more metal-rich}
than NGC~6441, as indicated by the presence of strong residuals for
several other metal lines, such as Ca4227, G4300, Fe4383 (indicated
in Figure~\ref{ratio}), Mg $b$, Fe5270, Fe5335 (not shown) and many
others.  Running data for both GCs through {\tt EZ\_Ages} \citep{gs08},
we find that [Fe/H] is approximately 0.2 dex higher in 225-280 than
in NGC~6441.  Therefore, it would not be at all unexpected that CN
bands are stronger in 225-280, even if the GC has the same abundance
pattern than that of NGC~6441.  When the spectrum of NGC~6441 is
divided by the average spectrum of GCs that are more similar in
metallicity, all residuals are considerably smaller (black line).
Comparison of the spectrum of NGC~6441 to those of some M~31 GCs
with similar metallicity, such as 073-134, 106-168, and 383-318,
shows essentially no residuals at the positions of CN bands.  We
note that the average spectrum used to obtain the ratio displayed
in Figure~\ref{ratio} includes GCs selected according to the following
criteria: $H\delta_F = H\delta_{F\,6441} \pm 0.3 {\rm\AA}$ and
$\langle{\rm Fe}\rangle, =\, \langle{\rm Fe}\rangle _{6441} \pm 0.2
{\rm\AA}$.  The reason for using $H\delta_F$ instead of $H\beta$
has to do with the presence of blue horizontal-branch stars in
NGC~6441.  This is going to be discussed in a lot more detail in a
forthcoming paper.

As a final note, we call attention to the fact that there
is a very strong residual in the CaII H and K lines at 3968 and
3933 ${\rm\AA}$, respectively.  We currently do not understand the
origin of this substantial difference between M~31 and MW GCs.
If real, this issue merits further scrutiny.  See discussion in
Section~\ref{thecase}.

In summary, we conclude that the data by \cite{be04} are also
consistent with our finding, that there is no substantial difference
between Galactic and M~31 GCs in terms of their CN strengths.  The
conclusion by \cite{be04} that the violet CN band indicates such
differences is possibly the by-product of a mismatch between the
metallicities of the GCs analyzed, combined with flux calibration
uncertainties in the violet part of the spectrum, where S/N is lower
and the sensitivity function as a function of wavelength in CCD-based
optical spectrographs is steep.

\subsection{The case of NGC~6528 and NGC~6553} \label{thecase}

Based on the previous section, one concludes that there is no
significant difference in CN strength between M~31 and MW GCs with
$[Fe/H] \simless -0.4$.  The same cannot be said about more metal-rich
GCs, since the three data points at the high-metallicity end of the
MW GC sample have substantially lower CN indices than their M~31
counterparts at the same [Fe/H].  Moreover, these three high-[Fe/H]
MW points seem to depart significantly from the
$\langle$Fe$\rangle$--$CN_1$ and $\langle$Fe$\rangle$--$CN_2$ trends
established by the MW GCs alone, by being displaced towards low CN
values.  The two highest $\langle$Fe$\rangle$ points are from two
spectra of NGC~6528, obtained with different spatial extraction
windows during data reduction \citep[see][for details]{s05}, whereas
the third point comes from the spectrum of NGC~6553 (a third
metal-rich GC, NGC~5927, also seems to be marginally too weak in
CN).  Visual confirmation of this result is offered at the bottom
panel of Figure~\ref{spectra} where the spectrum of NGC~6528 is
compared to the average spectrum of its M~31 counterparts.  CN bands
seem to be stronger in the average M~31 GC spectrum than in that
of NGC~6528.  Moreover, all features bluer than 4000 ${\rm\AA}$,
and particularly the CaII H and K lines, are weaker in NGC~6528.

The GCs NGC~6553 and 6528 are also the two most metal-rich in
the \cite{pu02} sample, and no such difference between these GCs
and their M~31 counterparts is seen in their data (Figure~\ref{puziadata}).
According to \cite{pu02}, the $CN_1$ index is stronger in the spectra
of NGC~6528 and 6553 by 0.035 and 0.096 mag than in \cite{s05}.
One therefore is left wondering whether these differences may be
due to different treatments of sky subtraction and spatial sampling
adopted by \cite{pu02} and \cite{s05}.  Let us recall that NGC~6528
and 6553 are located towards the Galactic bulge, in a region of the
sky that is affected by strong background and foreground contamination,
rendering the task of sky subtraction rather tricky and prone to
large uncertainties.  Moreover, these GCs, and NGC~6553 in
particular, are fairly sparse, making a fair representation of all
stellar types contributing relevantly to the integrated light rather
difficult, particularly for giant stars (regardless of the strengths
of CN bands in their spectra).

\begin{figure}
\includegraphics[angle=0,scale=0.8]{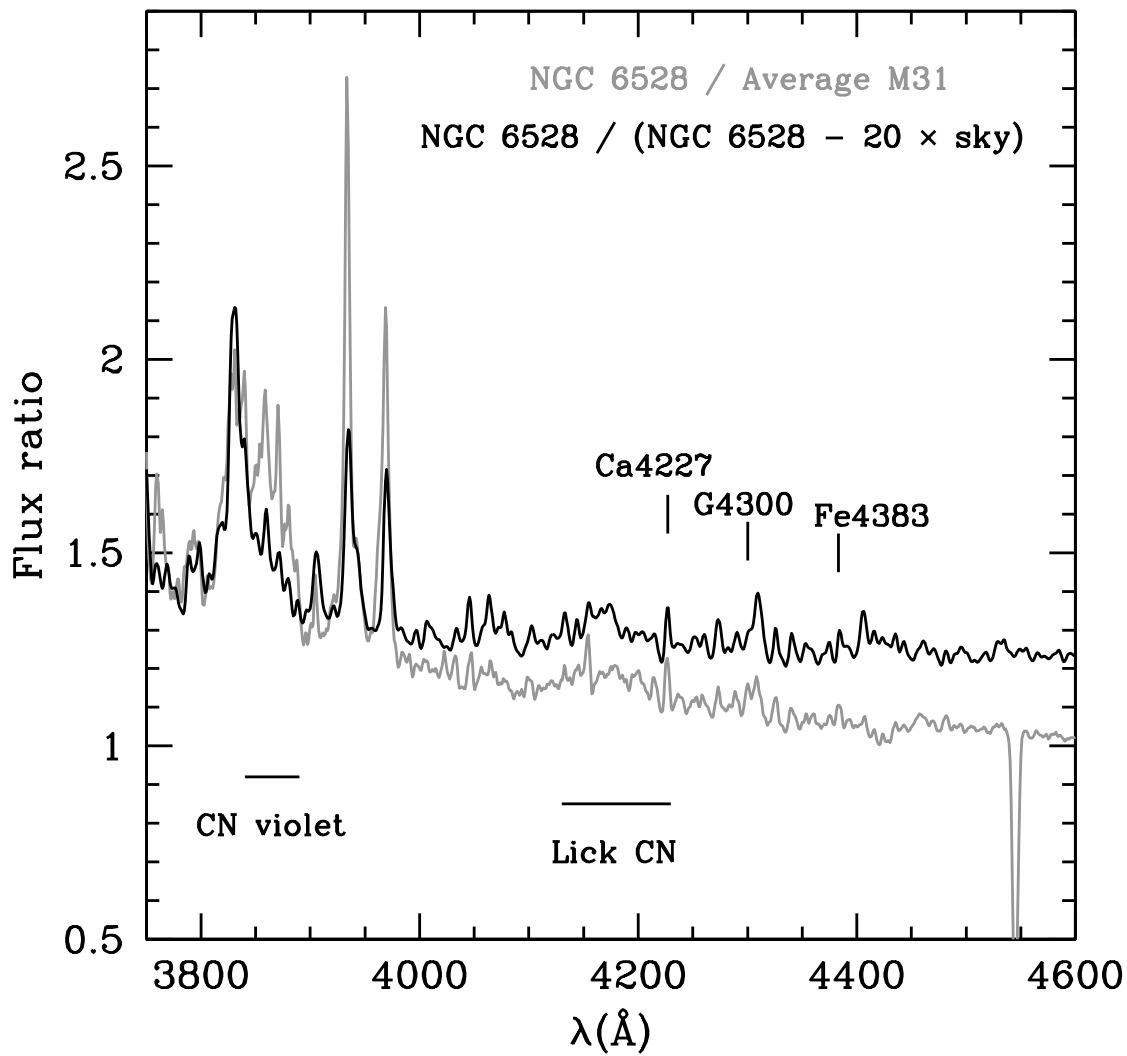}
\caption{Ratio between the spectra of NGC~6528 and M~31 GCs (gray
line), compared to a simulation of the effect of an error in the
sky subtraction (black line).  The latter is achieved by dividing
the spectrum of NGC~6528 by that same spectrum, subtracted of a sky
spectrum that is artificially boosted by a factor of 20.  The two
ratio spectra look somewhat similar, suggesting that the difference
between NGC~6528 and 6553 and their M~31 metal-rich counterparts
may be partly due to sky-subtraction uncertainties.  However, the
sky-subtraction error required to match the data is exceedingly
high (see text), and it nevertheless cannot match the residuals in
the Ca H and K lines.
}
\label{ratio_6528}
\end{figure}

A ratio between the spectra of NGC~6528 and its M~31 counterparts
from Table~\ref{lookalikes}, displayed in Figure~\ref{ratio_6528}
(gray line), shows the apparent signature of excess absorption in
the regions of the violet and Lick CN bands in spectra of M~31 GCs.
In order to check whether sky subtraction errors can be responsible
for the differences found, we generated a new NGC~6528 spectrum,
where the sky subtraction was changed by multiplying the sky level
by 20.  The ratio between NGC~6528 and that over-sky-subtracted
spectrum is shown as a solid line.  One can see that the spectra
look somewhat similar, suggesting that sky-subtraction errors may
be partly responsible for the discrepancies between metal-rich M~31
GCs and NGC~6528.  However, that would correspond to a huge error
in sky subtraction at $\lambda$ $\sim$ 4000 ${\rm\AA}$, which is
probably higher than the expected sky-background uncertainties.  We
note that, even in that case, the CN violet band in the ratio
spectrum is poorly matched by the simulated ratio, leading to the
conclusion that an even higher, and less likely, error in the sky
subtraction is required to explain the CN differences between
NGC~6528 and its M~31 counterparts.  Moreover, the Ca HK line
residuals in Figure~\ref{ratio_6528} are very discrepant, suggesting
that the difference in these lines cannot be explained by sky
subtraction errors alone.

In addition to differences in sky-subtraction procedures, the
\cite{pu02} spectra differ from those discussed in this work in the
way that the GC stars are sampled (Section~\ref{pudata}).  While
\cite{s05} drift-scanned a long slit within the GC core radii,
\cite{pu02} took a few spectra at fixed slit positions at varying
radial distance from the GC center.  Therefore, differences in the
way the two spectra sample the brightest GC stars can also lead to
measurable differences, particularly in line indices that tend to
be stronger in the spectra of giant stars.  The problem with this
explanation is that the stars that are subject to this kind of
stochastic effect contribute little light in the blue part of the
spectrum.

Finally, NGC~6528 and 6553 may possess an unusually high population
of blue stars, particularly blue stragglers.  We find this very
unlikely, since the residuals between these GC spectra and
their M~31 counterparts show no strong evidence for contamination
by A star light.  This result also argues against the presence of
any important contamination by foreground A stars.

If nevertheless background subtraction errors, and stochastic
effects, and hot stars in the foreground or in the GC are not to
blame, and indeed the GCs differ in their chemical composition, it
is not entirely clear whether the differences between Galactic
metal-rich clusters and their M~31 counterparts can be ascribed to
CN-enhancement of the latter.  The ratio spectrum suggests that
there may be a slight metallicity difference between NGC~6528 and
its M~31 counterparts selected on the basis of $H\beta$ and
$\langle$Fe$\rangle$
strengths, since residuals in other metal lines are also visible
in Figure~\ref{ratio_6528}, which may suggest the presence of
differences between the two GC samples in abundance ratios other
than [C/Fe] and [N/Fe].

Before concluding this section, we would like to comment on the
differences between the strengths of CaII HK in M~31 and MWGC
spectra.  One can clearly see in comparisons of spectra in
Figure~\ref{spectra} and in the ratio spectra showed in Figures~\ref{ratio}
and \ref{ratio_6528}, that CaII H and K lines are stronger in M~31
GC spectra than in those in the MW sample.  The differences are
more significant on the high metallicity end.  The hypothesis that
CN differences between NGC~6528 and 6553 and their counterparts in
M~31 are due to sky-subtraction errors could possibly also account
for these CaII H and K differences.  Because they are so strong,
the signal in the core of the line is fairly weak, and so these
lines are particularly sensitive to errors in background subtraction.
However, we note that a similar difference in CaII HK strength is
seen between the spectra of NGC~6441 and its M~31 counterparts by
\cite{be04}, on the basis of different datasets.  We compared CaII
HK strengths measured in the Hectospec spectra with those measured
in spectra obtained with Keck/LRIS by \citep{st10} and found them
to be consistent with each other, after accounting for the different
instrumental resolutions.  Therefore, we conclude that the Hectospec
spectra are free of important systematic effects due to background
subtraction.  Regarding the MW data, the similarity in the strengths
of the residuals in the Ca H and K lines in the ratio spectra of
Figure~\ref{ratio_6528} suggests that there indeed may be a background
subtraction problem with the spectra of these two metal-rich MW
GCs, but that cannot be the whole story, as the size of the
sky-subtraction error required to explain the data is too large---and
even in that case, it cannot explain differences in the Ca H and K
lines.  Independent data would help solving this puzzle.

To summarize, we are not entirely convinced that NGC~6528 and 6553
indeed differ from their M~31 counterparts in terms of their carbon
and nitrogen relative abundances, in spite of the fact that CN bands
seem to be weaker in their spectra than in M~31 metal-rich GCs.
It is possible that the differences are partly due to sky-subtraction
uncertainties, but this requires errors that seem unreasonably high.
The best way to approach this problem is through a study of CN
strengths in statistically significant samples of resolved stars
in those two bulge Milky Way GCs.  \cite{ms09} have recently
collected medium-resolution spectra of roughly a dozen individual
stars in NGC~6528 and found a few of them to be CN-strong.  Whether
that would translate into a CN-strong integrated spectrum or not,
is a question that can only be answered on the basis of spectra of
a much larger sample of GC members.

\section{Summary} \label{epilogue}

We present absorption line index measurements taken in integrated
spectra of a large number of M~31 and MW GCs, from \cite{ca09a}
and \cite{s05}, respectively.  We discuss the conversion of
instrumental measurements to a common equivalent system \citep[defined
in][]{s07}, as well as the uncertainties in the zero points of these
systems.

By comparing measurements of CN indices in old GCs belonging
to both data sets, we conclude that, in disagreement with previous
work by \cite{bu84}, \cite{bh90}, \cite{da90}, \cite{bu04},
\cite{be04}, and \cite{pu05}, among others, M~31 and MW 
GCs of the same [Fe/H] {\it generally} have similar CN-band
strengths.  Because this result disagrees with conclusions by many
different groups, we have reanalyzed the data by some of these
different authors and suggest that in most cases their conclusions
were a result of calibration problems.  In particular, the blue
Lick CN indices are relatively weak and are very sensitive to flux
calibration uncertainties.

We also find that the two MW GCs with [Fe/H] $\simgreater
-0.4$, NGC~6528 and 6553, have weaker CN bands than their M~31
counterparts at the same [Fe/H].  It is not entirely clear whether
this latter difference is real or a by-product of sky-subtraction
errors, or of issues with the sampling of the GCs' brightest
stars.  Other data sets \citep[e.g.,][]{pu05} do not show similar
CN discrepancies between these GCs and their M~31 metal-rich
counterparts.  Even if our results are correct, and these two
GCs indeed have weaker CN indices than metal-rich M~31 GCs,
the fact that they seem to depart significantly from a $\langle$Fe$\rangle$--$CN_1$
trend that is present in {\it both} samples suggests that NGC~6528
and 6553 are the exception rather than the rule.


The result by \cite{bu84} that M~31 GCs have stronger CN bands than
their MW counterparts of same metallicity, even though confirmed
by several later studies, has always been difficult to understand
because it is very hard to explain how nitrogen abundances could
differ substantially between otherwise very similar systems.  \cite{lb03}
proposed a scenario where GCs were formed from zero-metallicity
material, pre-enriched by hypernovae explosions in the center of
$\sim 10^6 M_\odot$ gas clouds.  Nitrogen abundance differences
between the two GC systems was the one piece of evidence that could
not be explained in that scenario.  We hope that the task of devising
a mechanism for the formation of the haloes of the Milky Way and
Andromeda galaxies is made easier by the findings presented in this
paper.  In a forthcoming paper, we contrast measurements of the
abundance patterns of M~31 and Galactic GCs, and discuss their
implications to our understanding of the formation of the two galaxy
haloes.

\acknowledgments 

We thank Sandy Faber for her insightful comments on an early version
of this manuscript and Jay Strader for inspiring discussions.  We
also acknowledge the fundamental contribution made by Jim Rose at
the early stages of this project.  An anonymous referee is thanked
for useful suggestions deriving from a careful and thorough reading
of the original manuscript.  RPS appreciates the support from Gemini
Observatory, which is operated by the Association of Universities
for Research in Astronomy, Inc., on behalf of the international
Gemini partnership of Argentina, Australia, Brazil, Canada, Chile,
the United Kingdom, and the United States of America.  The hospitality
of the Department of Astrophysical Sciences at Princeton University,
where this paper was partly conceived, is warmly acknowledged.  S.C.
acknowledges the support through a Discovery Grant from the Natural
Sciences and Engineering Research Council of Canada.

\appendix

\begin{deluxetable}{crrrrrrrrrrrrrrr} 
\rotate
\tabletypesize{\scriptsize}
\tablecaption{Zero-point conversions between the equivalent width
system of the Tololo spectra of MW globular clusters and the Lick indices from
	\cite{s07}}
\tablewidth{0pt}
\tablehead{\colhead{Index} & \colhead{$H\delta_A$} & \colhead{$H\delta_F$} &
           \colhead{CN$_1$} & \colhead{CN$_2$} & \colhead{Ca4227} & 
           \colhead{G4300} & \colhead{$H\gamma_A$} & \colhead{$H\gamma_F$} &
           \colhead{Fe4383} & \colhead{C$_2$4668} & \colhead{$H\beta$} & 
           \colhead{Mg$_2$} & \colhead{Mg $b$} & 
           \colhead{Fe5270} & \colhead{Fe5335} }
\startdata
Zero point$^1$ (${\rm\AA}$) &  +0.08 & --0.03 & --0.0088 & --0.0065 & --0.03 & --0.28 & +0.45 &
              +0.10 & --0.09 & +0.28 & --0.04   & --0.019 & --0.14 &
             --0.07  &  --0.09 \\
r.m.s.  (${\rm\AA}$)  &  0.25 &  0.12 &  0.011  &  0.0081 &  0.04  &  0.20 &  0.46 &
              0.22 &  0.27 & 0.13 &  0.08   &  0.015 &  0.15 &  0.14 &
              0.08  \\
\enddata
\tablenotetext{1}{Zero point = $I_{Tololo} - I_{Schiavon07}$} 
\label{zeromw}
\end{deluxetable}
\clearpage

\begin{deluxetable}{crrrrrrrrrrrrrrrr} 
\rotate
\tabletypesize{\scriptsize}
\tablecaption{Zero-point conversions between the equivalent width
system of the Hectospec spectra of M~31 clusters and the Lick indices from
	\cite{s07}}
\tablewidth{0pt}
\tablehead{\colhead{Index} & \colhead{$H\delta_A$} & \colhead{$H\delta_F$} &
           \colhead{CN$_1$} & \colhead{CN$_2$} & \colhead{Ca4227} & 
           \colhead{G4300} & \colhead{$H\gamma_A$} & \colhead{$H\gamma_F$} &
           \colhead{Fe4383} & \colhead{C$_2$4668} & \colhead{$H\beta$} & 
           \colhead{Fe5015} & \colhead{Mg$_2$} & \colhead{Mg $b$} & 
           \colhead{Fe5270} & \colhead{Fe5335} }
\startdata
Zero point$^1$ (${\rm\AA}$) & --0.25 & --0.06 & +0.0054 & +0.0054 & +0.01 & --0.01 & +0.01 &
             --0.01 & +0.12 & --0.04 & +0.04 & --0.03  & --0.0086 & --0.07 &
             +0.04  &  --0.04 \\
r.m.s.  (${\rm\AA}$)  &  0.26 &  0.10 &  0.008  &  0.011 &  0.10  &  0.13 &  0.18 &
              0.09 &  0.24 & 0.41 &  0.14 & 0.17  &  0.005 &  0.15 &  0.13 &
              0.11  \\
\enddata
\tablenotetext{1}{Zero point = $I_{Hectospec} - I_{Schiavon07}$} 
\label{zerom31}
\end{deluxetable}

\begin{deluxetable}{ccccc}
\tablecolumns{14}
\tablewidth{0pc}
\tablecaption{M~31 clusters used in the calculation of mean spectra for
	comparison with MW globular clusters}
\tablehead{\colhead{NGC~2298} & \colhead{NGC~1851} & 
	\colhead{47~Tuc}  & \colhead{NGC~6441} &
	\colhead{NGC~6528}}
\startdata
B003-G045  & B011-G063 & B001-G039 & B001-G039   & B055-G116    \\
B010-G062  & B027-G087 & B005-G052 & B004-G050   & B056-G117   \\
B022-G074  & B058-G119 & B016-G066 & B005-G052   & B143-G198   \\
B085-G147  & B076-G138 & B024-G082 & B008-G060   & B144-G000   \\
B165-G218  & B083-G146 & B034-G096 & B013-G065   & B147-G199   \\
B176-G227  & B148-G200 & B036-G000 & B016-G066   & B163-G217   \\
B188-G239  & B159-G000 & B037-G000 & B019-G072   & B193-G244   \\
B237-G299  & B161-G215 & B039-G101 & B034-G096   &   \\
\enddata
\label{lookalikes}
\end{deluxetable}

\begin{deluxetable}{lrrrrrrrrrrrrr}  
\rotate
\tabletypesize{\scriptsize}
\tablecolumns{14}
\tablewidth{0pc}
\tablecaption{Lick indices and uncertainties for standard stars observed with MMT/Hectospec}
\tablehead{\colhead{ID} &
\colhead{$H\delta_A$} & \colhead{$H\delta_F$}  & \colhead{$CN_1$} & \colhead{$CN_2$} &
\colhead{Ca4227} & \colhead{G4300}  & \colhead{$H\gamma_A$} & \colhead{$H\gamma_F$} &
\colhead{Fe4383} & \colhead{Ca4455}  & \colhead{Fe4531} & \colhead{$C_24668$} &
\colhead{$H\beta$}}
\startdata
HD~41597  &  -4.49  &   -1.13  &   0.136  &   0.165  &    0.58  &    6.96  &
-8.39  &   -2.93  &    4.50  &    1.29  &    3.37  &    4.23  &    1.38  \\ 
  &      0.06  &    0.04  &   0.001  &   0.002  &    0.03  &    0.04  &
  0.05  &    0.03  &    0.06  &    0.03  &    0.04  &    0.06  &    0.02
  \\ 
HD~41597  &  -4.57  &   -1.19  &   0.136  &   0.167  &    0.58  &    6.92  &
-8.31  &   -2.87  &    4.44  &    1.27  &    3.38  &    4.34  &    1.36  \\ 
  &      0.05  &    0.04  &   0.001  &   0.002  &    0.03  &    0.04  &
  0.05  &    0.03  &    0.05  &    0.03  &    0.04  &    0.06  &    0.02
  \\ 
HD~41597  &  -4.54  &   -1.05  &   0.140  &   0.170  &    0.56  &    6.85  &
-8.33  &   -2.89  &    4.46  &    1.26  &    3.44  &    4.47  &    1.36  \\ 
  &      0.06  &    0.04  &   0.001  &   0.002  &    0.03  &    0.04  &
  0.05  &    0.03  &    0.06  &    0.03  &    0.04  &    0.06  &    0.02
  \\ 
HD~43380  &  -6.23  &   -1.49  &   0.264  &   0.305  &    1.03  &    6.54  &
-9.33  &   -3.20  &    6.12  &    1.72  &    3.65  &    8.45  &    1.50  \\ 
  &      0.10  &    0.07  &   0.003  &   0.003  &    0.05  &    0.07  &
  0.08  &    0.05  &    0.09  &    0.05  &    0.07  &    0.09  &    0.04
  \\
\enddata
\label{std_m31a}
\end{deluxetable}
\clearpage

\begin{deluxetable}{lrrrrrrrrrrrr}  
\rotate
\tabletypesize{\scriptsize}
\tablecolumns{13}
\tablewidth{0pc}
\tablecaption{Lick indices and uncertainties for standard stars observed
	with MMT/Hectospec, continued}
\tablehead{\colhead{ID} &
\colhead{Fe5015} & \colhead{Mg$_1$}  & \colhead{Mg$_2$} & \colhead{Mg $b$} &
\colhead{Fe5270} & \colhead{Fe5335}  & \colhead{Fe5406} & \colhead{Fe5709} &
\colhead{Fe5782} & \colhead{$Na_D$}  & \colhead{$TiO_1$} & \colhead{$TiO_2$}}
\startdata
HD41597   &  5.11  &  0.0699  &  0.1441  &    2.19  &    2.78  &    2.28  &
1.51  &    1.02  &    0.68  &    1.58  &  0.0093  &  0.0181  \\
  &              0.05  &  0.0005  &  0.0006  &    0.02  &    0.02  &
  0.03  &    0.02  &    0.01  &    0.01  &    0.02  &  0.0004  &  0.0004
  \\ 
HD41597   &  5.10  &  0.0706  &  0.1430  &    2.18  &    2.80  &    2.22  &
1.52  &    1.03  &    0.68  &    1.58  &  0.0091  &  0.0174  \\ 
  &              0.04  &  0.0005  &  0.0005  &    0.02  &    0.02  &
  0.03  &    0.02  &    0.01  &    0.01  &    0.02  &  0.0004  &  0.0003
  \\ 
HD41597   &  5.10  &  0.0717  &  0.1442  &    2.19  &    2.76  &    2.26  &
1.51  &    0.99  &    0.67  &    1.56  &  0.0097  &  0.0185  \\ 
  &              0.05  &  0.0005  &  0.0006  &    0.02  &    0.02  &
  0.03  &    0.02  &    0.02  &    0.01  &    0.02  &  0.0004  &  0.0004
  \\ 
HD43380   &  6.08  &  0.1393  &  0.2583  &    3.83  &    3.57  &    3.08  &
2.13  &    1.31  &    1.10  &    2.61  &  0.0109  &  0.0309  \\ 
  &              0.07  &  0.0008  &  0.0009  &    0.04  &    0.04  &
  0.04  &    0.03  &    0.02  &    0.02  &    0.03  &  0.0007  &  0.0006
  \\
\enddata
\label{std_m31b}
\end{deluxetable}
\clearpage

\begin{deluxetable}{lrrrrrrrrrrrrr}  
\rotate
\tabletypesize{\scriptsize}
\tablecolumns{14}
\tablewidth{0pc}
\tablecaption{Lick indices for standard stars observed at CTIO}
\tablehead{\colhead{ID} &
\colhead{$H\delta_A$} & \colhead{$H\delta_F$}  & \colhead{$CN_1$} & \colhead{$CN_2$} &
\colhead{Ca4227} & \colhead{G4300}  & \colhead{$H\gamma_A$} & \colhead{$H\gamma_F$} &
\colhead{Fe4383} & \colhead{Ca4455}  & \colhead{Fe4531} & \colhead{$C_24668$} &
\colhead{$H\beta$}}
\startdata
HD~76151 &  -0.77  &    0.82  &  -0.030  &  -0.016  &    0.77  &    4.85  &
-3.85  &   -0.33  &    3.52  &    0.84  &    6.54  &    3.89  &    2.67  \\ 
HD~76924 &  -4.59  &   -1.31  &   0.196  &   0.224  &    0.40  &    6.28  &
-7.64  &   -2.20  &    4.80  &    1.30  &    7.16  &    5.51  &    1.58  \\ 
HD~78558 &  -0.50  &    0.92  &  -0.035  &  -0.022  &    0.63  &    5.01  &
-3.88  &   -0.37  &    3.36  &    0.76  &    6.37  &    3.62  &    2.55  \\ 
HD102574 &   1.79  &    1.92  &  -0.068  &  -0.051  &    0.59  &    3.95  &
-1.24  &    1.27  &    2.35  &    0.78  &    6.52  &    3.30  &    3.34  \\ 
HD157881 &  -2.45  &   -0.02  &  -0.088  &  -0.053  &    4.24  &    3.74  &
-7.76  &   -2.71  &    5.80  &    1.67  &    9.66  &    0.47  &   -0.82  \\ 
HD~76151 &  -0.68  &    0.91  &  -0.033  &  -0.017  &    0.73  &    4.78  &
-3.89  &   -0.35  &    3.58  &    0.82  &    6.83  &    3.79  &    2.67  \\ 
HD157881 &  -2.51  &   -0.05  &  -0.088  &  -0.053  &    4.25  &    4.01  &
-8.01  &   -2.93  &    5.63  &    1.66  &   10.16  &    0.58  &   -0.91  \\ 
\enddata
\label{std_mwa}
\end{deluxetable}

\begin{deluxetable}{lrrrrrrrrrrr}  
\rotate
\tabletypesize{\scriptsize}
\tablecolumns{13}
\tablewidth{0pc}
\tablecaption{Lick indices for standard stars observed at CTIO.
The Fe5015 index could not be measured in the CTIO sample, because
of the presence of a group of bad CCD columns crossing the spectra
within the index passband.}
\tablehead{\colhead{ID} &
\colhead{Mg$_1$}  & \colhead{Mg$_2$} & \colhead{Mg $b$} &
\colhead{Fe5270} & \colhead{Fe5335}  & \colhead{Fe5406} & \colhead{Fe5709} &
\colhead{Fe5782} & \colhead{$Na_D$}  & \colhead{$TiO_1$} & \colhead{$TiO_2$}}
\startdata
HD~76151 &  0.0191  &  0.1303  &    3.00  &    2.27  &    1.86  &
1.07  &    0.76  &    0.44  &    1.77  & -0.0429  & -0.0411  \\ 
HD~76924 &  0.0374  &  0.1246  &    2.09  &    2.90  &    2.34  &
1.62  &    1.11  &    0.75  &    1.96  & -0.0405  & -0.0122  \\ 
HD~78558 &  0.0209  &  0.1156  &    2.70  &    2.31  &    1.76  &
0.97  &    0.70  &    0.45  &    1.73  & -0.0465  & -0.0286  \\ 
HD102574 &  0.0074  &  0.0779  &    1.70  &    1.91  &    1.58  &
0.80  &    0.67  &    0.36  &    1.09  & -0.0461  & -0.0184  \\ 
HD157881 &  0.3540  &  0.4627  &    4.34  &    4.50  &    4.43  &
3.10  &    0.70  &    0.95  &   10.34  & -0.0164  &  0.0916  \\ 
HD~76151 &  0.0156  &  0.1271  &    3.05  &    2.25  &    1.86  &
1.00  &    0.73  &    0.42  &    1.63  & -0.0472  &  0.0082  \\ 
HD157881 &  0.3521  &  0.4670  &    4.41  &    4.56  &    4.46  &
3.10  &    0.74  &    0.98  &   10.35  & -0.0210  &  0.1036  \\ 
\enddata
\label{std_mwb}
\end{deluxetable}
\clearpage

\begin{deluxetable}{lrrrrrrrrrrrrr}
\rotate
\tabletypesize{\scriptsize}
\tablecolumns{14}
\tablewidth{0pc}
\tablecaption{Lick indices and uncertainties for Galactic globular clusters
	from \cite{s05}}
\tablehead{\colhead{ID} &
\colhead{$H\delta_A$} & \colhead{$H\delta_F$}  & \colhead{$CN_1$} & \colhead{$CN_2$} &
\colhead{Ca4227} & \colhead{G4300}  & \colhead{$H\gamma_A$} & \colhead{$H\gamma_F$} &
\colhead{Fe4383} & \colhead{Ca4455}  & \colhead{$C_24668$} & \colhead{$H\beta$}}
\startdata
NGC104  &   -0.48  &    0.68  &   0.025  &   0.050  &    0.56  &    4.66  &
-4.27  &   -0.66  &    2.49  &    0.67  &    1.68  &    1.60  \\ 
  &      0.02  &    0.02  &   0.001  &   0.001  &    0.01  &    0.02  &
  0.02  &    0.01  &    0.03  &    0.02  &    0.03  &    0.01  \\ 
NGC1851 &    2.23  &    2.06  &  -0.056  &  -0.031  &    0.34  &    2.79  &
-0.60  &    1.23  &    1.28  &    0.43  &    0.57  &    2.27  \\ 
  &      0.04  &    0.02  &   0.001  &   0.001  &    0.02  &    0.04  &
  0.03  &    0.02  &    0.05  &    0.03  &    0.06  &    0.02  \\ 
NGC1904 &    3.17  &    2.61  &  -0.085  &  -0.062  &    0.21  &    1.89  &
0.87  &    1.88  &    0.70  &    0.23  &   -0.07  &    2.48  \\ 
  &      0.06  &    0.04  &   0.002  &   0.002  &    0.04  &    0.06  &
  0.06  &    0.04  &    0.09  &    0.05  &    0.11  &    0.04  \\ 
NGC2298 &    3.70  &    2.87  &  -0.102  &  -0.079  &    0.24  &    1.66  &
1.45  &    2.13  &    0.65  &    0.15  &    0.17  &    2.68  \\ 
  &      0.09  &    0.07  &   0.003  &   0.004  &    0.06  &    0.10  &
  0.09  &    0.06  &    0.14  &    0.07  &    0.16  &    0.06  \\
\end{deluxetable}
%
%
NGC6528$^1$ &   -1.36  &    0.37  &   0.059  &   0.088  &    0.88  &    4.81  &
-5.75  &   -1.27  &    4.58  &    1.16  &    4.41  &    1.59  \\ 
  &      0.21  &    0.15  &   0.006  &   0.007  &    0.11  &    0.17  &
  0.19  &    0.11  &    0.23  &    0.12  &    0.24  &    0.09  \\ 
NGC6528$^2$ &   -1.45  &    0.30  &   0.063  &   0.093  &    0.89  &    4.82  &
-5.79  &   -1.31  &    4.67  &    1.17  &    4.51  &    1.57  \\ 
  &      0.17  &    0.12  &   0.005  &   0.006  &    0.09  &    0.14  &
  0.15  &    0.09  &    0.19  &    0.10  &    0.19  &    0.08  \\ 
NGC6544 &    1.85  &    1.49  &  -0.056  &  -0.038  &    0.34  &    2.90  &
-1.48  &    0.53  &    1.47  &    0.40  &    0.33  &    1.60  \\ 
  &      0.19  &    0.13  &   0.005  &   0.006  &    0.10  &    0.16  &
  0.17  &    0.11  &    0.24  &    0.12  &    0.26  &    0.10  \\ 
NGC6553 &   -0.91  &    0.33  &   0.042  &   0.071  &    0.69  &    4.68  &
-5.11  &   -1.03  &    3.87  &    0.92  &    3.95  &    1.56  \\ 
  &      0.16  &    0.11  &   0.004  &   0.005  &    0.08  &    0.13  &
  0.14  &    0.09  &    0.19  &    0.10  &    0.20  &    0.07  \\
\enddata
\tablenotetext{1}{Measured on the spectrum derived from the smaller
	extraction aperture.  See details in \cite{s05}}
\tablenotetext{2}{Measured on the spectrum derived from the larger
	extraction aperture.  See details in \cite{s05}}
\end{deluxetable}
\clearpage

\begin{deluxetable}{lrrrrrrrrrrr}
\rotate
\tabletypesize{\scriptsize}
\tablecolumns{13}
\tablewidth{0pc}
\tablecaption{Lick indices and uncertainties for Galactic globular clusters
	from \cite{s05}, continued.}
\tablehead{\colhead{ID} &
\colhead{Mg$_1$}  & \colhead{Mg$_2$} & \colhead{Mg $b$} &
\colhead{Fe5270} & \colhead{Fe5335}  & \colhead{Fe5406} & \colhead{Fe5709} &
\colhead{Fe5782} & \colhead{$Na_D$}  & \colhead{$TiO_1$} & \colhead{$TiO_2$}}
\startdata
NGC~104  &  0.0569  &  0.1589  &    2.69  &    1.91  &    1.68  &    1.06
&    0.64  &    0.46  &    1.57  &  0.0203  &  0.0581  \\ 
  &            0.0003  &  0.0004  &    0.01  &    0.02  &    0.02  &
  0.01  &    0.01  &    0.01  &    0.01  &  0.0003  &  0.0003  \\ 
NGC1851 &  0.0251  &  0.0829  &    1.35  &    1.40  &    1.12  &    0.67  &
0.46  &    0.22  &    1.21  & -0.0169  &  0.0046  \\ 
  &            0.0005  &  0.0006  &    0.02  &    0.03  &    0.03  &
  0.02  &    0.02  &    0.02  &    0.02  &  0.0005  &  0.0005  \\ 
NGC1904 &  0.0103  &  0.0485  &    0.83  &    0.93  &    0.78  &    0.34  &
0.22  &    0.12  &    0.90  & -0.0111  & -0.0003  \\ 
  &            0.0009  &  0.0011  &    0.05  &    0.05  &    0.06  &
  0.04  &    0.04  &    0.03  &    0.05  &  0.0010  &  0.0009  \\ 
NGC2298 &  0.0070  &  0.0442  &    0.79  &    0.74  &    0.63  &    0.24  &
0.18  &    0.18  &    1.96  & -0.0122  &  0.0005  \\ 
  &            0.0013  &  0.0016  &    0.07  &    0.07  &    0.08  &
  0.06  &    0.05  &    0.05  &    0.07  &  0.0014  &  0.0013  \\
\end{deluxetable}
%
%
NGC6528$^1$ &  0.1041  &  0.2515  &    3.80  &    2.77  &    2.42  &    1.67  &
1.03  &    0.80  &    4.96  &  0.0313  &  0.1118  \\ 
  &            0.0019  &  0.0023  &    0.09  &    0.09  &    0.10  &
  0.08  &    0.06  &    0.06  &    0.07  &  0.0016  &  0.0015  \\ 
NGC6528$^2$ &  0.1025  &  0.2489  &    3.76  &    2.80  &    2.42  &    1.69  &
1.01  &    0.80  &    5.00  &  0.0324  &  0.1087  \\ 
  &            0.0015  &  0.0019  &    0.07  &    0.08  &    0.09  &
  0.06  &    0.05  &    0.05  &    0.06  &  0.0014  &  0.0012  \\ 
NGC6544 &  0.0413  &  0.0934  &    1.00  &    1.37  &    1.07  &    0.65  &
0.55  &    0.63  &    2.14  & -0.0079  & -0.0053  \\ 
  &            0.0020  &  0.0024  &    0.09  &    0.10  &    0.12  &
  0.09  &    0.07  &    0.06  &    0.08  &  0.0018  &  0.0016  \\ 
NGC6553 &  0.0944  &  0.2420  &    3.82  &    2.74  &    2.20  &    1.68  &
1.01  &    1.13  &    3.59  &  0.0231  &  0.0900  \\ 
  &            0.0015  &  0.0018  &    0.07  &    0.08  &    0.09  &
  0.06  &    0.05  &    0.05  &    0.06  &  0.0014  &  0.0012  \\
\enddata
\tablenotetext{1}{Measured on the spectrum derived from the smaller
	extraction aperture.  See details in \cite{s05}}
\tablenotetext{2}{Measured on the spectrum derived from the larger
	extraction aperture.  See details in \cite{s05}}
\end{deluxetable}
\clearpage

\begin{deluxetable}{lrrrrrrrrrrrrrrr}
\rotate
\tabletypesize{\scriptsize}
\tablecolumns{14}
\tablewidth{0pc}
\tablecaption{Lick indices for M~31 Globular clusters.  Sample only.  The
	full table is available in the electronic version.}
\tablehead{\colhead{ID} &
\colhead{$CN_1$} & \colhead{$CN_2$} &
\colhead{Ca4227} & \colhead{G4300}  & 
\colhead{Fe4383} & \colhead{Ca4455}  & \colhead{Fe4531} &
\colhead{$C_24668$} & \colhead{$H\beta$} &
\colhead{Fe5015} & \colhead{Mg$_1$}}
\startdata
B001-G039 &   0.003  &   0.024  &    0.65  &    4.92  &    2.95  &    0.80
&    2.48  &    2.06  &    1.68  &    3.48  &  0.0522  \\
  &     0.008  &   0.010  &    0.15  &    0.23  &    0.34  &    0.16  &
  0.23  &    0.32  &    0.12  &    0.24  &  0.0023   \\
B002-G043 &  -0.089  &  -0.073  &    0.05  &    0.82  &    0.13  &    0.07
&    0.66  &    0.18  &    2.55  &    0.53  &  0.0099  \\
  &     0.004  &   0.006  &    0.09  &    0.15  &    0.24  &    0.11  &
  0.17  &    0.24  &    0.09  &    0.20  &  0.0020   \\
B003-G045 &  -0.075  &  -0.057  &    0.25  &    1.74  &    0.71  &    0.19
&    1.31  &    0.30  &    2.68  &    1.33  &  0.0161  \\
  &     0.005  &   0.007  &    0.10  &    0.17  &    0.29  &    0.13  &
  0.19  &    0.28  &    0.11  &    0.23  &  0.0023   \\
B004-G050 &   0.014  &   0.035  &    0.56  &    4.14  &    2.55  &    0.76
&    2.63  &    2.45  &    1.78  &    3.47  &  0.0478  \\
  &     0.005  &   0.006  &    0.09  &    0.14  &    0.22  &    0.11  &
  0.15  &    0.22  &    0.09  &    0.18  &  0.0018   \\
B005-G052 &   0.042  &   0.066  &    0.65  &    4.47  &    2.43  &    0.51
&    2.51  &    2.10  &    1.48  &    3.69  &  0.0561  \\
  &     0.005  &   0.006  &    0.10  &    0.15  &    0.23  &    0.12  &
  0.17  &    0.25  &    0.10  &    0.21  &  0.0021   \\
\enddata
\end{deluxetable}

\begin{deluxetable}{lrrrrrrrrrr}
\rotate
\tabletypesize{\scriptsize}
\tablecolumns{13}
\tablewidth{0pc}
\tablecaption{Lick indices for M~31 Globular Clusters.  Sample only.  The
	full table is available in the electronic version.}
\tablehead{\colhead{ID} & 
\colhead{Mg$_2$} & \colhead{Mg $b$} &
\colhead{Fe5270} & \colhead{Fe5335}  & \colhead{Fe5406} & \colhead{Fe5709} &
\colhead{Fe5782} & \colhead{$Na_D$}  & \colhead{$TiO_1$} & \colhead{$TiO_2$}}
\startdata
B001-G039 &  0.1604  &    2.63  &    2.02  &    1.68  &    1.21  &    0.69
&    0.65  &    3.31  &  0.0262  &  0.0509  \\
  &            0.0027  &    0.11  &    0.12  &    0.13  &    0.10  &
  0.07  &    0.07  &    0.09  &  0.0020  &  0.0016  \\
B002-G043 &  0.0410  &    0.62  &    0.44  &    0.59  &    0.37  &    0.08
&    0.14  &    1.02  &  0.0094  & -0.0029  \\
  &            0.0023  &    0.10  &    0.11  &    0.13  &    0.09  &
  0.08  &    0.08  &    0.12  &  0.0022  &  0.0018  \\
B003-G045 &  0.0588  &    0.91  &    0.87  &    0.77  &    0.52  &    0.41
&    0.24  &    1.49  &  0.0106  & -0.0038  \\
  &            0.0026  &    0.11  &    0.12  &    0.14  &    0.10  &
  0.09  &    0.08  &    0.13  &  0.0025  &  0.0020  \\
B004-G050 &  0.1480  &    2.27  &    2.04  &    1.66  &    1.13  &    0.67
&    0.41  &    2.06  &  0.0232  &  0.0372  \\
  &            0.0021  &    0.08  &    0.09  &    0.11  &    0.08  &
  0.06  &    0.06  &    0.09  &  0.0018  &  0.0015  \\
B005-G052 &  0.1590  &    2.51  &    2.11  &    1.64  &    1.11  &    0.70
&    0.54  &    2.94  &  0.0237  &  0.0374  \\
  &            0.0025  &    0.10  &    0.11  &    0.12  &    0.09  &
  0.07  &    0.07  &    0.09  &  0.0020  &  0.0017  \\
\enddata
\end{deluxetable}
\clearpage

\begin{deluxetable}{lrrrrrrrr}
\rotate
\tabletypesize{\scriptsize}
\tablecolumns{13}
\tablewidth{0pc}
\tablecaption{Conversion between Tololo and Indo-US Lick/IDS index
	systems, for indices not modeled by \cite{s07}, with associated
		uncertainties.}
\tablehead{\colhead{Index} & \colhead{Ca4455} & 
\colhead{Mg$_1$}  & \colhead{Fe5406} & \colhead{Fe5709} &
\colhead{Fe5782} & \colhead{NaD}     & \colhead{$TiO_1$} & \colhead{$TiO_2$}}
\startdata
Conversion$^1$ & -0.05 & -0.01 & -0.02 & -0.02 & -0.029 &  0.10 & -0.006 & -0.02 \\
r.m.s.         &  0.03 & 0.02 & 0.04 &  0.02 &  0.008 & 0.07 & 0.004 & 0.03 \\
\enddata
\tablenotetext{1}{Zero point = $I_{Tololo} - I_{Indo-US}$} 
\label{convmw}
\end{deluxetable}

\begin{deluxetable}{lrrrrrrrrr}
\rotate
\tabletypesize{\scriptsize}
\tablecolumns{13}
\tablewidth{0pc}
\tablecaption{Conversion between Hectospec and Indo-US Lick/IDS index
	systems, for indices not modeled by \cite{s07}, with associated
		uncertainties.}
\tablehead{\colhead{Index} & \colhead{Ca4455} &  \colhead{Fe4531} &
\colhead{Mg$_1$}  & \colhead{Fe5406} & \colhead{Fe5709} &
\colhead{Fe5782} & \colhead{NaD}     & \colhead{$TiO_1$} & \colhead{$TiO_2$}}
\startdata
Conversion$^1$ & 0.03 & -0.11 & 0.02 & -0.06 & -0.03 & -0.004 & -0.03 &
-0.002 & -0.002 \\
r.m.s.         & 0.07 &  0.21 & 0.01 & 0.09 &  0.07 &  0.036 & 0.13 & 0.005
& 0.005 \\
\enddata
\tablenotetext{1}{Zero point = $I_{Hectospec} - I_{Indo-US}$} 
\label{convm31}
\end{deluxetable}

\begin{deluxetable}{lrrrrrrrrr}
\rotate
\tabletypesize{\scriptsize}
\tablecolumns{13}
\tablewidth{0pc}
\tablecaption{Conversion between Indo-US and standard Lick/IDS index
	systems, for indices not modeled by \cite{s07}, with associated
		uncertainties.}
\tablehead{\colhead{Index} & \colhead{Ca4455} & \colhead{Fe4531} &
\colhead{Mg$_1$}  & \colhead{Fe5406} & \colhead{Fe5709} &
\colhead{Fe5782} & \colhead{NaD}     & \colhead{$TiO_1$} & \colhead{$TiO_2$}}
\startdata
Conversion$^1$ & -0.28 & -0.05 & -0.009 & 0.08 & -0.03 & -0.03 & -0.08 &
-0.0015 & 0.003 \\
r.m.s.         &  0.25 &  0.31 & 0.022 & 0.35 &  0.17 &  0.13 & 0.25 &
0.0084 & 0.012 \\
\enddata
\tablenotetext{1}{Zero point = $I_{Indo-US} - I_{Worthey94}$} 
\label{conv}
\end{deluxetable}

{}

\clearpage









\clearpage


\begin{thebibliography}{}
\bibitem[Barmby \etal (2000)]{ba90} Barmby, P., Huchra, J., Brodie, J.,
	Forbes, D>, Schroder, L. \& Grillmair, C. 2000, \aj, 119, 727
\bibitem[Beasley \etal (2004)]{be04} Beasley, M.A., Brodie, J.P., Strader,
	J., Forbes, D.A., Proctor, R.N., Barmby, P. \& Huchra, J.P. 2004,
	\aj, 128, 1623
\bibitem[Brodie \& Hanes (1986)]{bh86} Brodie, J.P. \& Hanes, D. 1986, 
	\apj, 300, 258
\bibitem[Brodie \& Huchra (1990)]{bh90} Brodie, J.P. \& Huchra, J.P. 1990,
	\apj, 362, 503
\bibitem[Brodie \& Huchra (1991)]{bh91} Brodie, J.P. \& Huchra, J.P. 1991,
	\apj, 379, 157
\bibitem[Burstein \etal (1984)]{bu84} Burstein, D., Faber, S.M., Gaskell,
	C.M., \& Krumm, N. 1984, \apj, 287, 586
\bibitem[Burstein \etal (2004)]{bu04} Burstein, D., Li, Y., Freeman, K.C.
	\etal 2004, \apj, 614, 158
\bibitem[Caldwell \etal\ (2009)]{ca09a} Caldwell, N., Harding, P.,
	Morrison, H., Rose, J.A., Schiavon, R. \& Kriessler, J. 2009a, \aj,
	137, 94
\bibitem[Caldwell \etal\ (2011)]{ca10} Caldwell, N., Schiavon, R.P.,
        Morrison, H., Rose, J.A. \& Harding, P. \aj, 141, 61
\bibitem[Cannon \etal\ (1998)]{can98} Cannon, R.D., Croke, B.F.W., Bell,
	R.A., Hesser, J.E. \& Stathakis, R.A. 1998, \mnras, 298, 601
\bibitem[Cardiel \etal\ (1998)]{ca98} Cardiel, N., Gorgas, J., Cenarro, J.
\& Gonz\'alez, J.J. 1998, \aaps, 127, 597
\bibitem[Carretta \& Gratton (1997)]{cg97} Carretta, E. \& Gratton, R.
	1997, \aaps, 121, 95
\bibitem[Carretta \etal (2009)]{ca09} Carretta, E., Bragaglia, A., Gratton,
	R., D'Orazi, V. \& Lucatello, S. 2009, \aap, 508, 695
\bibitem[Catelan (2009)]{cat09} Catelan, M. 2009, \apss, 320, 261
\bibitem[Cohen \etal (1998)]{co98} Cohen, J.G., Blakeslee, J.P. \& Ryzhov,
	A. 1998, \apj, 496, 808
\bibitem[Colucci \etal\ (2009)]{co09} Colucci, J.E., Bernstein, R.A.,
	Cameron, S., McWilliam, A. \& Cohen, J.G. 2009, \apj, 704, 385
\bibitem[Conroy \& Spergel (2010)]{cs10} Conroy, C. \& Spergel, D.N. 2010,
	\apj, submitted, {\tt arXiv: 1005.4934v1}
\bibitem[Davidge (1990)]{da90} Davidge, T.J. 1990, \apj, 351, L37
\bibitem[Dorman (1992)]{do92} Dorman, B. 1992, \apjs, 81, 221
\bibitem[Fabricant \etal (2005)]{fa05} Fabricant, D. \etal 2005, \pasp,
	117, 1411
\bibitem[Fan \etal (2008)]{fa08} Fan, Z., Ma, J., de Grijs, R. \& Zhou, X.
	2008, \mnras, 385, 1973
\bibitem[Freitas Pacheco \& Barbuy (1995)]{fb95} Freitas Pacheco, J.A. \&
Barbuy, B. 1995, \aap, 302, 718
\bibitem[Gratton \etal (2004)]{gr04} Gratton, R., Snedden, C. \& Carretta,
	E. 2004, \araa, 42, 385
\bibitem[Graves \& Schiavon (2008)]{gs08} Graves, G.J. \& Schiavon, R.P. 2008,
      \apjs, 177, 446.
\bibitem[Gray (2008)]{gr08} Gray, D.F. 2008, The Observation and Analysis
of Stellar Photospheres (3rd ed; Cambridge: Cambridge Univ. Press)
\bibitem[Jones (1999)]{jo99} Jones, L.A. 1999, Ph.D. Thesis, Univ. North
Carolina
\bibitem[Harris (1996)]{ha96} Harris, W.E. 1996, \aj, 112, 1487
\bibitem[Kraft \& Ivans (2003)]{ki03} Kraft, R.P. \& Ivans, I.I. 2003,
	\pasp, 115, 143
\bibitem[Lee \etal (2000)]{le00} Lee, H.-C., Yoon, S.-J. \& Lee, Y.-W.
2000, \aj, 120, 998
\bibitem[Li \& Burstein (2003)]{lb03} Li, Y \& Burstein, D. 2003, \apj,
	614, L29
\bibitem[Martell \& Smith (2009)]{ms09} Martell, S.L. \& Smith, G.H. 2009,
	\pasp, 121, 577
\bibitem[Morrison \etal (2010)]{mo10} Morrison, H., Caldwell, N.C.,
	Schiavon, R.P., Athanassoula, L., Harding, P. \& Rose, J.A. 2010,
	\apj, submitted.
\bibitem[Ocvirk (2010)]{oc10} Ocvirk, P. 2010, \apj, 709, 88
\bibitem[Perrett \etal (2002)]{pe02} Perrett, K.M., Bridges, T.J., Hanes,
	D.A., \etal 2002, \aj, 123, 2490
\bibitem[Piotto (2009)]{pi09} Piotto, B. 2009, IAUS, 258, 233
\bibitem[Piotto \etal\ (2002)]{pi02} Piotto, G. \etal\ 2002, \aap\, 391,
	945
\bibitem[Ponder \etal (1998)]{po98} Ponder, J.M., Burstein, D., O'Connell,
	R.W., \etal 1998, \aj, 116, 2297
\bibitem[Poole \etal (2010)]{po10} Poole, V., Worthey, G., Lee, H.-L. \&
	Serven, J. 2010, \aj, 139, 809
\bibitem[Prochaska \etal (2007)]{pr07} Prochaska, L.C., Rose, J.A.,
	Caldwell, N., Castilho, B.V., Concannon, K., Harding, P., Morrison,
	H. \& Schiavon, R.P.  2007, \aj, 134, 231
\bibitem[Puzia \etal (2002)]{pu02} Puzia, T.H., Saglia, R.P.,
	Kissler-Patig, M., Maraston, C., Greggio, L., Renzini, A. \&
	Ortolani, S. 2002, \aap, 395, 45
\bibitem[Puzia \etal (2005)]{pu05} Puzia, T.H., Perrett, K.M. \& Bridges,
	T.J. 2005, \aap, 434, 909
\bibitem[Rey \etal (2007)]{re07} Rey, S.C. \etal 2007, apjs, 173, 643
\bibitem[Rich \etal (1997)]{ri97} Rich, R.M., Sosin, C., Djorgovski, S.G.,
	Piotto, G., King, I.R., Renzini, A., Phinney, E.S., Dorman, B.,
	Liebert, J. \& Meylan, G.  1997, \apj, 484, L25
\bibitem[Rose (1994)]{ro94} Rose, J.A. 1994, \aj, 107, 206
\bibitem[Salasnich \etal\ (2000)]{sa00} Salasnich, B., Girardi, L., Weiss,
	A. \& Chiosi, C. 2000, \aap, 361, 1023
\bibitem[S\'anchez-Bl\'azquez \etal (2003)]{sb03} S\'anchez-Bl\'azquez, P.,
	Gorgas, J., Cardiel, N., Cenarro, J., Gonz\'alez, J.J. 2003, \apj,
	590, L91
\bibitem[Sandage \& Wallerstein (1960)]{sw60} Sandage, A. \& Wallerstein,
	G. 1960, \apj, 131, 598
\bibitem[Schiavon (2007)]{s07} Schiavon, R.P. 2007, \apjs, 171, 146
\bibitem[Schiavon \etal (2002)]{s02} Schiavon, R.P., Faber, S.M., Castilho,
	B.V. \& Rose, J.A. 2002, \apj, 580, 850
\bibitem[Schiavon \etal\ (2004a)]{s04a} Schiavon, R.P., Caldwell, N. \& Rose,
	J.A. 2004, \aj, 127, 1513.
\bibitem[Schiavon \etal\ (2004b)]{s04b} Schiavon, R.P., Rose, J.A., Courteau,
	S. \& MacArthur, L.A. 2004b, \apj, 608, L33
\bibitem[Schiavon \etal\ (2005)]{s05} Schiavon, R.P., Rose, J.A., Courteau,
	S. \& MacArthur, L.A. 2005, \apjs, 160, 163
\bibitem[Schiavon \etal\ (2011)]{s11} Schiavon, R.P. \etal 2011, \aj, in preparation.
\bibitem[Strader (2010, priv. comm.)]{st10} Strader, J. 2010, private
communication.
\bibitem[Trager \etal (1998)]{tr98} Trager, S.C., Worthey, G., Faber, S.M.,
	Burstein, D. \& Gonz\'alez, J.J. 1998, \apjs, 116, 1
\bibitem[Trager \etal (2005)]{tr05} Trager, S.C., Worthey, G., Faber, S.M.
\& Dressler, A.  2005, \mnras, 362, 2
\bibitem[Valdes \etal (2004)]{va04} Valdes, F., Gupta, R., Rose, J.A.,
	Singh, H.P. \& Bell, D.J. 2004, \apjs, 152, 251
\bibitem[Wallace (1962)]{wa62} Wallace, L. 1962, \apjs, 7, 165
\bibitem[Worthey \etal\ (1994)]{wo94} Worthey, G., Faber, S.M., Gonz\'alez,
J.J. \& Burstein, D. 1994, \apjs, 94, 687
\bibitem[Worthey \& Ottaviani (1997)]{wo97} Worthey, G. \& Ottaviani, D.L. 
1997, ApJS, 111, 377
\bibitem[Yoon \etal\ (2006)]{yo06} Yoon, S.-J, Yi, S.K. \& Lee, Y.-W. 2006,
	Science, 311, 1129

\end{thebibliography}
\end{document}